\documentclass[prd,aps,nofootinbib,preprintnumbers, preprint, 12pt]{revtex4}
\usepackage{amsmath,amssymb,epsf}
\usepackage{graphicx, pstricks}

\def\be{\begin{equation}}
\def\ee{\end{equation}}
\def\ba{\begin{eqnarray}}
\def\ea{\end{eqnarray}}

\def\lta{~\mbox{\raisebox{-.6ex}{$\stackrel{<}{\sim}$}}~}
\def\gta{~\mbox{\raisebox{-.6ex}{$\stackrel{>}{\sim}$}}~}
\def\bq{\begin{quote}}
\def\eq{\end{quote}}

\def\Mp{M_{\mathrm{Pl}}}

\begin{document}


\title{Dangerous Angular KK/Glueball Relics in String Theory Cosmology}

\author{J.F. Dufaux$^{1,2}$, L. Kofman$^{2}$, M. Peloso$^{3}$}

\affiliation{$^1$ Instituto de F\'isica Te\'orica \ UAM/CSIC,
Universidad Aut\'onoma de Madrid, Cantoblanco, 28049 Madrid, Spain}
\affiliation{$^2$ CITA, University of Toronto, 60 St. George st.,
Toronto, ON M5S 3H8, Canada}
\affiliation{$^3$ School of Physics and Astronomy, University of
Minnesota, Minneapolis, MN 55455, USA}

\date{\today}

\begin{abstract}

{The presence of Kaluza-Klein particles in the universe is a potential manifestation of
string theory cosmology. In general, they can be present in  the high temperature bath of the early universe.
In particular examples,  string theory inflation 
often ends with brane-antibrane annihilation followed by the energy cascading 
through massive closed string loops to  KK modes which then decay into  lighter
standard model particles. 
However, massive KK modes in the early universe may  become  dangerous cosmological relics if the inner manifold
 contains warped throat(s) with approximate isometries.
In the complimentary picture, in the AdS/CFT dual gauge theory with extra isometries, massive glueballs of
various spins become  the dangerous cosmological relics.
 The decay of these angular KK modes/glueballs,  located around the tip of the throat,
 is caused  by   isometry breaking which results from gluing the throat    to the compact CY  manifold.
 We address  the problem  of these angular KK particles/glueballs, studying  their interactions and decay channels,
from the theory side,  and
the resulting cosmological constraints on the warped compactification parameters,
from the phenomenology side. The abundance and decay time of the long-lived
non-relativistic angular KK modes depend strongly on the parameters of the warped 
geometry, so that observational constraints rule out  a   significant fraction
of the parameter space. In particular, the coupling of the angular KK particles can be weaker than gravitational.}

\end{abstract}

\preprint{IFT-UAM/CSIC-07-66}
\preprint{UMN-TH-2637/08}
\maketitle

\section{Introduction}

Often when we embed a new layer of elementary particle theory in the theory of
the early universe, we find that new particles become dangerous cosmological relics. 
The study of this problem is one of the few tools available for theorists to bridge 
these two theories. For example, the GUT theory incorporated in the Hot Big Bang model 
predicted the overproduction of monopoles; SUGRA theory predicted the overproduction of 
gravitinos and moduli fields. It is interesting to explore if the embedding of string theory 
in cosmology results in new dangerous relics.

Kaluza-Klein (KK) particles in the universe is a  possible signature  of
string theory physics. Relic KK particles may  be good or bad for cosmology, depending on their properties. 
If their abundance multiplied by their mass is tuned to be right, and their life time is 
significantly longer than the age of the universe, the KK particles can be dark matter candidates. 
Otherwise, and more typically, they are dangerous relics, which may overclose the universe, or violate 
 Big Bang Nucleosynthesis or other astrophysical constraints, depending on their abundance and lifetime.
Therefore, the theory of relic KK particles is a powerful tool to study string theory cosmology and to confront 
it with observations. 

The AdS/CFT duality allows us to translate from the language of the gravitational KK sector
to gauge theories. Therefore, independently of string theory, the story of dangerous cosmological 
relics  repeats itself in the context of gauge theories with extra isometries, which contain
massive glueballs of different spins. 

We will consider KK particles in a  string theory and string theory cosmology setting,
which drew significant interest in the last few years, and which can be advanced quantitatively. 
Indeed, important progress has been made  in the studying   type IIB string theory 
compactifications with fluxes where the moduli are stabilized~\cite{GKP, KKLT}.
These developments in string theory have raised interests in the embedding of early universe inflation 
and low energy physics in a fundamental and phenomenologically viable framework. 

Constructions with fluxes and conifolds induce warped regions  of spacetime
\cite{GKP,KT,KS} - throats - which in turn may be used to generate mass  hierarchies.
For instance, the hierarchy between the string scale and the electroweak scale may be generated 
\`{a} la Randall-Sundrum~\cite{RS} when the Standard Model degrees of freedom are localized 
at the bottom of a sufficiently warped throat.
Furthermore,  warping is an essential ingredient in the attempts to realize
brane inflation in the slow-roll regime~\cite{K2LM2T} or the DBI 
regime~\cite{DBI}, or in the fast-roll regime~\cite{KM07}, see also 
e.g.~\cite{Baumann, Krause:2007jk, Becker:2007ui} for recent works. 

Warped brane/anti-brane inflation involves a pair of brane/anti-brane $D3$ and $\bar D3$, which are 
string theory objects: although inflation in this model may be described with an effective four dimensional field theory,
its end point, reheating, is associated with the $D3-\bar D3$ annihilation, which is not captured by 
field theory. Reheating from brane/anti-brane annihilation is rather different from (p)reheating
in quantum field theory \cite{KLS}, and was investigated in several 
papers~\cite{Barnaby, lev, Frey:2005jk, Chialva:2005zy, tye2},
which  identified  the channels of inflaton energy decay, cascading from tachyon annihilation 
to the lighter standard model particles through massive closed string loops,
 KK modes, and brane displacement moduli. 
Indeed, if the internal space has warped regions (throats), the mass of the KK modes associated with them is much lower
 than the fundamental scale, being redshifted by 
the warp factor at the bottom of the throat. If the KK modes
are sufficiently light then  it is natural to expect them 
to be produced during reheating.
For instance, in models of brane/anti-brane inflation, KK modes are copiously produced 
by the decay of the closed string loops from the brane/anti-brane annihilation at the end of inflation, and their
 energy must be efficiently converted into Standard Model (SM) degrees of freedom.

Notably, reheating is then bottlenecked through KK modes which are localized around the tip of the throat. 
Warped brane inflation and reheating after brane annihilation were  considered for throat models 
with high isometries, often 
described by the Klebanov-Strassler (KS) solution \cite{KS}. The KS geometry has $S0(4)\times \mathbb{Z}_2$ 
isometries around the tip of the throat, and it is close to $AdS_5\times T^{1,1}$ away from  the tip.
Other known examples of throat solutions also  have similar isometries \cite{Butti:2004pk}.
To make the inner space compact, the KS throat is glued to the unwarped Calabi-Yau (CY) part of the manifold.
For cosmologists, a significant advantage of this model is that the metric of the throat is known analytically,
so that KK modes can be studied in some detail. 
However, as it  was found in ~\cite{lev}, the advantage in mathematics is in this case a disadvantage in physics. Indeed, the
KK modes in internal spaces with isometries have conserved angular momentum quantum numbers associated with these isometries.
The final decay products of the KK particles, the SM particles, do not have such quantum numbers.
Therefore KK modes with (internal) angular momenta cannot decay into SM particles. Moreover,  KK particles with opposite angular
momenta cannot annihilate completely into SM particles due to the expansion of the universe. Therefore, we encounter the severe
 problem that the freezed out fraction of the long-lived angular Kaluza-Klein modes can overclose the universe~\cite{lev}
\footnote{It turns out that the left-over abundance of angular KK modes is very sensitive to the parameters of the 
throat geometry. Alternatively, after tuning the parameters, these modes have been 
proposed as possible dark matter candidates~\cite{lev}, \cite{tye2} (see also \cite{Servant:2002aq} for earlier proposals 
of KK dark matter, in the context of other models).}!
The problem of angular KK modes from an inner geometry with throats is not specific to 
the warped brane/anti-brane inflationary scenario.
Independently on the particular realization of inflation, light KK modes may be produced during reheating
if they are sufficiently coupled to inflation. 
In any context where the reheat temperature of the very early universe
is large enough, massive KK particles can be produced in the thermal bath. Then, KK modes with
 angular momenta will again freeze out, and can be dangerous for cosmology.

In a sense, the problem of dangerous angular KK modes in string theory cosmology is a 
re-appearance of the problem of KK modes from dimensional reduction in supegravity in higher dimensions
 with isometries \cite{Kolb:1983fm}. 
However, the string theory setting brings new features in cosmology with KK modes.
 The unwarped CY part of the compact manifold  does not
preserve the isometries of the throat. Therefore, when we embed  the KS throat into the compact manifold,
 the isometries of the throat inevitably get broken.
These deviations from isometries  are significant at the base of the throat, close to the unwarped CY, 
but they decrease exponentially towards the tip of the throat. On the other hand,  the KK modes are localized at the tip 
of the throat and their wave functions decrease exponentially towards the base. The net effect will be some violation of the 
conservation of the angular momentum of the KK modes. In terms of the four-dimensional effective lagrangian, 
this manifests itself by additional terms describing the decay of angular KK modes into  SM particles with 
small coupling constants, controlled by the isometry breaking. 
This makes angular KK modes in the universe not stable but short 
or long lived, depending on the parameters of the model.

In this paper, we investigate in some detail the decay of Kaluza-Klein modes with angular momenta 
associated with the compact manifold in warped flux compactifications, namely, in a KS throat with broken isometries. 
In Section \ref{sec:background} we begin with the background throat geometry; we consider first the KS solution with intact 
isometries, and we then introduce the isometry breaking. We use the static perturbations of the KS metric considered in
 \cite{Aharony:2005ez}, where they were connected to symmetry breaking operators in the context of the AdS/CFT correspondence.

In Section \ref{sec:modes} we consider the wave functions of the KK modes in the throat geometry. For definiteness we
 will discuss the gravi-tensor KK modes (the spin-two KK modes, or spin-two glueballs). The knowledge of the wave functions 
allows us to integrate
the internal space out, and to study the effective four-dimensional physics of the model. From the spectroscopy of 
the ``KK atoms'' (i.e. their masses as a function of the quantum numbers associated with the internal dimensions) we can constrain 
the interactions of the KK modes with each others and with SM particles; this results in ``selection rules'' which control 
the decay of these modes. We focus on the $AdS$ part of the geometry as an
approximation of the actual KS throat. 
We employ standard (quantum mechanics) perturbation theory to compute the effect of the isometry breaking perturbations 
on the wave functions of the KK modes.

The interactions of the KK modes are studied in Section \ref{sec:interactions}. We consider the  model 
where the SM fields are  localized on a  $3$-brane around  the tip of the same throat where the excited KK modes are located. 
We study the possible decay channels of the long-lived modes and find that the leading one is into SM particles.
Note that this channel differs from  the decay of scalar KK modes 
into massless gravi-vector particles which was  recently considered in \cite{Berndsen:2007my}.

The second part of our paper is devoted to the consequences that the KK modes can have for cosmology and astrophysics.
We first, in Section \ref{sec:abundance}, estimate the abundance of  angular KK modes which  freeze out in the
hot expanding  universe, and also estimate  their life-times. We will mostly work with the  conservative assumption of
 initial thermal equilibrium  of KK modes.

It turns out that the theoretical predictions for the angular KK modes abundance and life-time are very strongly dependent 
on the parameters of the modes. Therefore, in Section \ref{sec:constraints} we put together observational constraints on
the parameters of KK modes and consequently, on the underlying throat model, using limits from BBN \cite{moroi}
and the astrophysical $\gamma$-ray background \cite{diffusegamma}, in case of decaying modes, and limit 
on the dark matter energy density, in case of lifetimes much greater than the present age of the universe.

We restrict our calculations of the angular KK modes to single-throat settings. There are several 
reasons for that. The case of two throats, with one of them associated with (brane-antibrane) inflation
(of energy scale $10^{13}$ GeV or so)
and the other one associated with the standard model sector (of TeV energy scale), is often considered in the literature.
In the cosmological context, however, the story of KK modes in the two throat models with such significant
energy offset has the following conceptual complication, which usually is not addressed in the literature.
Indeed, first, the Hubble parameter scale $H\sim 10^{10}$GeV of inflation exceeds the TeV mass of 
the SM throat KK modes. Similarly, the temperature $T$ after brane-antibrane reheating exceeds that low mass scale.
For those situations the low energy SM throat will be screened by a horizon
\footnote{The emergence of the horizon in the KS throat was considered, e.g. in \cite{horizon}.}.
In other words, the
KK mass spectrum of the low energy throat has a gap below defined by $ H $ or $ T$.
Therefore, the simple picture of KK modes tunneling from the inflationary throat to the SM throat throat is 
relevant only after the temperature of the universe drops below the TeV scale,  
i.e. when the timing of tunneling is longer than $10^{-12}$ sec.
Another reason is that tunneling of the angular KK modes between the throats through the bulk of the inner manifold
is exponentially suppressed.

In the Discussion Section we summarize our calculations and discuss generalizations of our results 
to other SM models and multiple-throat cases.

\section{Background Warped Geometry} \label{sec:background}

In this section, we discuss the background geometry with the warped throat
as a part of the inner space manifold.   
We will use this construction  in the rest of the paper. 

We consider a type IIB compactification
with fluxes turned  along the internal dimensions.
A common  feature of flux compactification is warping.
The 10-D metric can be written 
\be
ds^2 = H^{-1/2}(z)  \,g_{\mu\nu}\,dx^{\mu} dx^{\nu} +  H^{1/2}(z)   {g}_{ab}(z)\,dz^a dz^b \ ,
\ee
or alternatively 
\be
\label{back}
ds^2 = e^{2A(y)} \,g_{\mu\nu}\,dx^{\mu} dx^{\nu} +  \hat{g}_{ab}(y)\,dy^a dy^b \ .
\ee
The greek indices $\mu, \nu, \lambda, ...$ run over the 4-dimensional coordinates $x^{\mu}$
of the external spacetime, while the latin indices $a, b, c, ...$ run over the 
6-dimensional coordinates $y^a$ of the internal compact space (whose metric is $\hat{g}_{ab}$). 
The outer space  of the 10-dimensional background is approximated with the 4-dimensional Minkowski metric, 
$g_{\mu\nu} = \eta_{\mu\nu}$
\footnote{This approximation is poor when the Hubble parameter of the expanding universe exceeds the typical mass scale 
at the tip of the throat (like the masses of KK modes). In this case the tip of the throat would be screened from the rest
of the compact manifold by an horizon and the KK mass spectrum would acquire a mass gap  $ \gta H$. 
In this paper $H \ll  m \sim T$. }.

The 10-dimensional geometry involves one or several ``throats'', each of them 
on one side ends with a regular ``tip'', and on the other side is smoothly 
glued to the bulk of the CY manifold \footnote{
The embedding geometry of the inner manifold  reminds us of ``an udder'' with multiple tips sticking out.
Baroque ``udder'' constructions are not so baroque for relativistic  astrophysics: recall that the embedding diagram
of the multiple supermassive black holes in the galaxies around us in an expanding universe also looks
like an udder!}. The known solutions for the throat, e.g. the KS solution, 
involve specific isometries and formally have a non-compact radial dimension.
In order  to get a sensible four-dimensional effective theory, the throat has to be glued to a compact 
CY which does not preserve these isometries. This results in deviations from the isometric KS solution,
which are increasing as we approach the bulk of the CY. On the other hand, these isometry breaking perturbations 
are expected to be exponentially suppressed at the bottom of the throat. 

In the following sub-section, we discuss in more details the geometry of the throat.
In sub-section \ref{isbr}, we consider the breaking of the isometries due to the compact CY. 
In sub-section \ref{param}, we discuss the essential parameters of the model.

\subsection{Isometric KS  throat}
\label{isth}

The known example of analytical throat solution ending with a regular tip 
is given by the Klebanov-Strassler solution \cite{KS}, where the warped throat contains a deformed conifold. 
The deformed conifold~\cite{Candelas} is a 6-dimensional space defined in 
$\mathbb{C}^4 = \mathbb{R}^8$ by the quadric
\be
\label{quadric}
\sum_{i=1}^4 z_i^2 = \varepsilon^2 \ ,
\ee
where $z_i$ are four complex coordinates. Far from the origin $z_i = 0$, the deformation 
$\varepsilon$ may be neglected and the space approaches a cone. Its base, given by the 
intersection of (\ref{quadric}) with the sphere in $\mathbb{C}^4$,
\be
\label{inter}
\sum_{i=1}^4 |z_i|^2 = \mathrm{constant} \ ,
\ee
is the 5-dimensional manifold $T^{1,1}$, a fibration of $U(1)$ over $S^2 \times S^2$, 
which has the topology of $S^2 \times S^3$.

The warping is achieved by turning on $M$ units of RR flux through a 3-cycle 
of the deformed conifold and $-K$ units of NS-NS flux through the dual 3-cycle.
The full KS solution \cite{KS} involves rather tedious formula for the metric. 
Not all of its information will be needed for our discussion, therefore we will
work with simpler approximations.
Away from the bottom of the throat, the geometry is approximately given by the 
Klebanov-Tseytlin solution~\cite{KT}
\be
\label{kt1}
ds^2 = H^{-1/2}(r)\,\eta_{\mu\nu}\,dx^{\mu} dx^{\nu} 
+ H^{1/2}(r)\,\left(dr^2 + r^2\,ds_5^2\right) \ ,
\ee
where $ds_5^2$ is the line element of $T^{1,1}$. The warping may be written
\be
\label{kt2}
H(r) = \frac{R_+^4 + R_-^4\,\ln\left(\frac{r}{R_+}\right)}{r^4}
\ee
with 
\be
\label{R+-}
R_+^4 = \frac{27\pi}{4}\,\alpha'^2\,g_s\,M\,K \;\;\;\;\; 
\mbox{ and } \;\;\;\;\; R_-^4 = \frac{81}{8}\,\alpha'^2\,g_s^2\,M^2 \ ,
\ee
where $\sqrt{\alpha'}$ is the string length, $g_s$ is the string coupling,
and the amounts of fluxes $M$ and $K$ are chosen such that $R_+ > R_-$ 
(see Eq. (\ref{yt}) below). This approximate solution is valid only 
in the range $R_+\,e^{-R_+^4/R_-^4 + 1} < r < R_+$. At smaller $r$, the 
singular metric based on the conifold has to be replaced by the regular one 
based on the deformed conifold~\cite{KS}. 
At $r \sim R_+$, the geometry is glued to the compact CY. 

Locally, the metric (\ref{kt1}) corresponds to the direct product 
of $AdS_5$ and $T^{1,1}$ with the same radius of curvature $R$, but with $R^4 = R_+^4 + R_-^4\,\ln\left(\frac{r}{R_+}\right)$ 
slowly (logarithmically) 
varying from $R = R_-$ to $R = R_+$. 
Taking $R$ to be a constant between $R_-$ 
and $R_+$, the metric reduces to the simple form 
\be
\label{isoback}
ds^2 = e^{-2y/R}\,\eta_{\mu\nu}\,dx^\mu dx^\nu + dy^2 
+ R^2\,f_{ij}(\Omega_5)\,d\theta^i d\theta^j \ ,
\ee
where the coordinate $y = R\,\ln(R/r)$ measures the proper distance 
in the radial direction, and goes from $y = 0$ to $y = y_t$. 
$f_{ij}$ is the metric on a 5-dimensional manifold $X^5$, approximated 
by $T^{1,1}$ in the present case, and $i, j, k, ...$ run over the 5 ``angular'' coordinates $\theta^i$ 
(denoted collectively as $\Omega_5$).

At the bottom of the KS throat (corresponding to $y > y_t$ in (\ref{isoback})), the 
$T^{1,1}$ evolves into a round $S^3$ of finite radius, while the two other angular 
directions, corresponding to the $S^2$ fibered over the $S^3$, shrink to zero size. 
The geometry may be described in terms of a new radial coordinate $\rho$, with 
$r \propto e^{\rho/3}$ at large $\rho$. 
The asymptotic behavior of the metric in the vicinity of the tip at $\rho = 0$ 
reads as 
\be
ds^2 = e^{-2y_t/R}\,\eta_{\mu\nu}\,dx^{\mu} dx^{\nu} + 
R_-^2\,\left[dS_3^2 + d\rho^2 + \rho^2\,e_{ij}(\theta^k)\,d\theta^i d\theta^j\right] \ ,
\ee 
where the warp factor $e^{-2 y_t/R}$ and the radius $R_-$ of $S^3$ are approximately 
constant, and $e_{ij}(\theta^k)$ denotes the metric of the $S^2$ fibered over the $S^3$. 
The total variation of the warp factor from the top to the bottom of the throat is 
approximately given by~\cite{GKP}
\be
\label{yt}
e^{-2 y_t/R} \simeq e^{-\frac{2\pi K}{3 g_s M}} \ ,
\ee 
which can generate naturally large hierarchies for suitable choices of fluxes.

The isometry group of $T^{1,1}$ is $SU(2) \times SU(2) \times U(1)$, 
as may be seen  from its embedding (\ref{inter}) and (\ref{quadric}) with $\varepsilon = 0$,
 where  $SU(2) \times SU(2) \simeq SO(4)$ rotates the $z_i$'s and $U(1)$ multiply them by a 
phase. This is broken into $SO(4) \times \mathbb{Z}_2$ by the deformation $\varepsilon$ 
of the conifold, where $SO(4)$ corresponds to rotations on the $S^3$. These are the 
isometries of the KS throat.

\subsection{Breaking Isometries by Gluing to CY}
\label{isbr}

Kaluza-Klein modes in the $4+6$ dimensional KS geometry with isometries in the angular 
dimensions carry angular momenta quantum numbers associated with them. 
Standard Model particles in the 4-dimensional outer spacetime do not have those
quantum numbers. Therefore in an expanding universe excited angular KK modes 
will never be converted completely into the SM particles. In the model with isometric
throat angular KK particles are the dangerous cosmological relics.

However, when the throat is glued to the  compact CY at  $y=0$, its geometry deviates more and more from 
(\ref{isoback}) as $y$ decreases towards zero. 
In this sub-section we consider the background  geometry of the throat which is smoothly glued to the big compact manifiold.
We will model this geometry as static perturbations around the isometric geometry (\ref{isoback}). 
We write \footnote{In Eq. (\ref{perback}), the 
perturbations $\epsilon(y)\,w(\Omega_5)$ and $\epsilon(y)\,\delta f_{ij}(\Omega_5)$ should be understood as sums over 
perturbations with different radial profiles, see Eq. (\ref{series}) and the discussion below. We will sometimes keep the 
``symbolic form'' $\epsilon(y)\,w(\Omega_5)$ and $\epsilon(y)\,\delta f_{ij}(\Omega_5)$ to simplify notations.} 
\be
\label{perback}
ds^2 = e^{-2y/R}\,\left[1 + \epsilon(y)\,w(\Omega_5)\right]\,\eta_{\mu\nu}\,dx^\mu dx^\nu + dy^2 + 
R^2\,\left[f_{ij}(\Omega_5) + \epsilon(y)\,\delta f_{ij}(\Omega_5)\right]\,d\theta^i d\theta^i \ ,
\ee
where $\delta f_{ij}(\Omega_5)$ and $w(\Omega_5)$ are the isometry breaking perturbations at a given $y$ of the metric 
on $X_5$ and of the warp factor respectively. For generic perturbations, 
only the 4-dimensional Lorentz invariance remains in (\ref{perback}). The perturbations are expected to be of order one 
at the base of the throat, where it is glued to the bulk of the CY, and to decrease exponentially 
away from it, $\epsilon(y) \sim e^{-\alpha y / R}$. 
In general, other supergravity fields and other components of the 10-dimensional metric may be perturbed as well, however, 
the form (\ref{perback}) will capture the effects we are interested in.

In principle, for a given throat model, the form of the isometry breaking perturbations may be calculated 
by harmonic analysis. The breaking of the $SO(4)$ isometry at the tip of the KS throat has been studied 
in~\cite{Aharony:2005ez} (see also \cite{DeWolfe:2004qx}), where it was interpreted in the language of 
the AdS/CFT correspondence. From the point of view of the field theory dual to the warped background, the gluing 
of the compact CY to the throat may be described by the addition of symmetry breaking operators in the UV. 
The RG flow of these operators corresponds to the radial profile of the perturbations of the 10-dimensional background.
On an $AdS_5 \times X^5$ background, a perturbation $\epsilon(y)\,\delta f_{ij}(\Omega_5)$ is decomposed into 
harmonics $\delta f_{ij}^{(L)}(\Omega_5)$ on $X^5$, each with its own profile in the radial direction
\be
\label{series}
\epsilon(y)\,\delta f_{ij}(\Omega_5) \to \sum_{\{L\}} 
e^{-\alpha_L y / R} \, \delta f^{(L)}_{ij}(\Omega_5) \ ,
\ee
where we have neglected the sub-leading contributions with respect to $y$. The radial profile follows 
from the linearized radial mode equation after dimensional reduction on $X^5$. The parameters 
$\alpha_L$ thus depend on the mode under consideration and its quantum numbers $L$ on $X^5$.
The leading perturbation in the throat arises from the mode with the lowest value of $\alpha_L$.

Ref.~\cite{Aharony:2005ez} classified the perturbations, breaking the $SO(4)$ isometry but preserving 4-dimensional 
Lorentz invariance and supersymmetry, which may be turned  on a KS throat approximated by $AdS_5 \times T^{1,1}$. 
The leading perturbation found in \cite{Aharony:2005ez} corresponds to a perturbation 
$e^{-\alpha_L y / R} \, \delta f^{(L)}_{ij}(\Omega_5)$ of the metric with 
$\alpha_L = \sqrt{28} - 4 \simeq 1.29$. It involves a tensor mode $\delta f^{(L)}_{ij}(\Omega_5)$ 
on $T^{1,1}$ with quantum numbers $L = (l_1 = 1, l_2 = 1, r = 0)$, where $(l_1, l_2, r)$ are the principal 
quantum numbers associated to the $SU(2)_1 \times SU(2)_2 \times U(1)$ isometry of $T^{1,1}$
(see \cite{Ceresole:1999ht} for the KK spectroscopy of the type IIB supergravity fields on $AdS_5 \times T^{1,1}$).
The isometry breaking perturbations involving other harmonics are exponentially suppressed in the throat compared to the leading one.  
Note also that, for a perturbation with given radial profile $e^{-\alpha_L y / R}$, 
the principal quantum numbers are fixed but the other ones (those which are not involved in the eigenvalues of the Laplacian 
on $X^5$) may vary since they do not appear in the radial mode equation.

Finally, Ref.~\cite{Aharony:2005ez} estimated the perturbation of the warp factor $e^{2A}$ at the tip of the KS throat 
induced by the perturbation $\delta f_{ij}$. The leading correction was found to be of the order 
$\delta A(y_t) \sim e^{-\alpha y_t / R}$ with the same coefficient $\alpha \simeq 1.29$. This mode is absent on the 
$AdS_5 \times T^{1,1}$ background but should correspond to a linear perturbation around the full KS solution. Again, 
we expect different harmonics in the angular directions to have different exponential suppressions at the tip of the throat. 
Because these modes arise from deviations of the KS solution from $AdS_5 \times T^{1,1}$, their radial profile in the throat 
may differ from $e^{-\alpha y / R}$. However, what will be important for us is their suppression at the tip, 
$\delta A(y_t) \sim e^{-\alpha y_t / R}$, and it will be convenient to keep the profile $e^{-\alpha y / R}$ in the 
throat to model this suppression. The perturbation of the metric in the angular directions and the perturbation of the 
warp factor will play similar roles in the following.

\subsection{Parameters of the Warped Geometry}
\label{param}

Here we discuss the basic parameters of the warped compactification, which will enter into
the formulas for the cosmological abundance and decay time of KK modes.
The cosmological constraints from the dangerous KK relics will be given in terms of these
parameters.

One of the essential parameters will be the leading exponent $\alpha_L$ (for the KK
modes with quantum numbers $L$) for the isometry breaking, entering in (\ref{perback}), (\ref{series}).
In addition to the parameters $\alpha$, and to the string length $\sqrt{\alpha'}$ and coupling $g_s$,
our treatment will involve essentially three other parameters:  
the (AdS) radius of curvature $R$ in (\ref{isoback}), the volume of the whole 6-dimensional 
internal space $V_6$, and the warp factor at the bottom of the throat $e^{-y_t/R}$. 
We now recall  useful relations between these parameters.

The volume $V_6$ is dominated by  the ``big'' compact CY manifold, and it is always 
greater than the contribution of the throat alone,  $V_6^{1/6} > R$. 
For instance, for the KS  throat this implies
$V_6 \, > \, \int_{\mathrm{throat}}d^6y \sqrt{\hat{g}} \, 
\simeq \, \mathrm{Vol}(T^{1,1}) R_+^6 / 2 \, = \, 8\pi^3 R_+^6 / 27$.  We will also take 
$R >  \sqrt{\alpha'}$ in order to work in  the supergravity approximation. 
In KS, $R \, > \, R_- \, > \, \sqrt{\alpha'}$ may be achieved by a suitable choice of fluxes, 
see Eq. (\ref{R+-}). 

The string length and $V_6$ are related through the value of the 4-dimensional Planck mass
\be
\label{Ms}
\Mp^2 \,  = \, \frac{2\,V_6}{(2\pi)^7\,g_s^2\,\alpha'^4} \ .
\ee 
We will take $g_s = 0.1$ in the estimations below.

The final parameter is the value of the warp factor at the bottom of the throat, $e^{-y_t/R}$.
We will distinguish the short throat and the long throat, when this parameter is
relatively big or relatively small, correspondingly. 
The  short throat is associated with the warped brane inflation proposed in~\cite{K2LM2T},
and is often called the inflationary throat.
In the original model, the normalisation 
of the adiabatic density perturbations generated during inflation reads as 
\be
\label{deltaH}
\delta_{H} \, \simeq \, 0.4 \, N_e^{5/6} \, 
\left(\frac{T_3}{\Mp^4}\right)^{1/3}\,e^{-4y_t / 3R} \, \simeq \, 2 \times 10^{-5}
\ee
at a number of e-foldings $N_e \approx 60$. Here $T_3 = \left[(2\pi)^3 g_s \alpha'^2\right]^{-1}$ 
is the tension of the $D3$, $\bar D3$ branes responsible for inflation. 
The original KKLMMT inflationary construction  has  been generalized by several models (e.g. 
\cite{DBI, Baumann, Krause:2007jk, Becker:2007ui}) for which the parameters may be different.  
Eq.~(\ref{deltaH}) will serve as a convenient reference value for a short throat.

The long throat is associated with the hierarchy between the string scale 
and the TeV (electroweak) scale \`a la Randall-Sundrum~\cite{RS}, if the Standard Model 
degrees of freedom are located on a brane around the tip of throat. In this case
the warping is significant  
\be
\label{hier}
e^{-y_t/R} \approx \frac{\mathrm{TeV}}{M_s} \ ,
\ee 
where $M_s = 1/\sqrt{\alpha'}$.
However, it does not mean that inflation cannot be associated with the long throat.
The original KKLMMT throat slow roll inflation suffers from the $\eta$-problem due to the conformal
coupling of the inflaton. It turns out that one still can obtain sufficient fast roll inflation with the conformal
coupling, but only in the long throat \cite{KM07}.
For this case, the estimation (\ref{deltaH}) for the amplitude of the inflaton cosmological fluctuations
is not applicable. 

Both Eqs. (\ref{hier}) and (\ref{deltaH}) determine the value of the following specific combination of the 
parameters
\begin{equation}
{\cal N} \equiv \left(\frac{V_6^{1/6}}{\sqrt{\alpha'}}\right)^3\;e^{y_t/R} \ .
\label{defN}
\end{equation}
We will thus use this combination in the calculations of the KK modes abundance and decay times.

For the short throat~(\ref{deltaH}), using Eq.~(\ref{Ms}) with $g_s = 0.1$, 
Eq.~(\ref{defN}) gives 
\be
\label{Ithroat}
{\cal N}  \approx \; 5 \times 10^5 
\hspace*{2cm} \mbox{( short throat)} \ .
\ee
In this case, significant warping ($e^{y_t/R} \gg 1$) requires at least $V_6^{1/6} / \sqrt{\alpha'} \ll 80$, 
and smaller values for $R / \sqrt{\alpha'}$ and $V_6^{1/6} / R$.
For  example, $V_6^{1/6} \approx 5\,\sqrt{\alpha'}$ leads to $e^{y_t/R} \approx 4\times 10^3$.

For the long throat, using Eqs.~(\ref{Ms}) and (\ref{hier}), we have 
\be
\label{SMthroat}
{\cal N}  \approx \; 10^{17} 
\hspace*{2cm} \mbox{(long  throat)} \ .
\ee

\section{Spin 2 Kaluza-Klein modes} \label{sec:modes} 

Let us recall the energy cascade at the end of the brane-antibrane inflation \cite{lev}.
Brane-antibrane annihilation through tachyon decay results in the production of massive
closed string loops, which further decay into  KK modes,
interacting with each others, and with  SM particles.
Massive KK modes are localized at the bottom of the throat, while massless four-dimensional gravitons are not.
There are various  KK modes of the inner manifold with the throat(s) with fluxes,
e.g. \cite{KK1,KK2}.
We will be interested in the KK modes with angular degrees of freedom associated with inner dimensions.
In the KK sector with angular momentum, bearing in mind cosmological applications, 
we shall be primarily concerned with the lightest modes localized at the bottom of the throat.
Unfortunately, the mass spectrum of KK modes in the KS geometry is not known exactly but only up to the order 
of magnitude estimation $m \sim e^{-y_t/R}/R$.

At this point the study can go into different directions, depending on the choice of the prototype KK relics.
Ref. \cite{Berndsen:2007my} considers a scalar KK mode, which is the lightest one on the $AdS_5 \times T^{1,1}$ 
background \footnote{Such a mode is not necessarily the lightest one after compactification (for this mode, 
specific boundary conditions have to be imposed on a slice of $AdS$, see \cite{Delgado:2003tx, Gherghetta:2000qt}) 
and after moduli stabilization.}. However, to make this mode unstable, \cite{Berndsen:2007my} introduced 
additional SUSY breaking operator, and the final decay products involve massless gravi-vector 
modes \footnote{In the third version of \cite{Berndsen:2007my}, the decay channel has been modified and the final 
decay products involve only massless gravitons.}. 
There can be other subtleties: next-to-the-lightest modes may be forbidden to decay into the lightest modes,
and their decay time directly into SM brane degrees of freedom may be different. We will encounter these
situations in our calculations below. Therefore, a comprehensive investigation shall include different possibilities.

In this paper, as a representative set of these KK modes, for economy and definiteness, 
we focus on the spin-2 fields in four-dimensions, 
and their decay into excitations of the SM brane located in the same throat. The dynamics of the spin-2 
KK modes follows from the ten-dimensional linearized Einstein equation.
Their wave equation depends only on the 10-dimensional background metric and is decoupled from the fluctuations 
of the matter sources. 
Our purpose is to calculate the abundance and lifetime of angular KK modes and make connection
between cosmological constraints and parameters of the model.
The methodology we develop in this paper can be extended for different KK modes and different
phenomenological settings of their decays. 

The spin-2 fields correspond to the symmetric transverse-traceless perturbation $h_{\mu\nu}$
\be
\label{spin2}
g_{\mu \nu} = \eta_{\mu\nu} + h_{\mu\nu}(x^{\lambda}, y^c) \;\;\;\;\; \mbox{ with }  
\;\;\;\;\; \eta^{\mu\lambda} \partial_{\lambda} h_{\mu\nu} = \eta^{\mu \nu} h_{\mu \nu}=0 \ .
\ee 
of the 10-dimensional metric (\ref{back}).

The equations of motion for the KK modes (\ref{spin2}) in the background geometry (\ref{back}) are derived in the 
Appendix. The 4-dimensional Poincare invariance allows to solve by separation of 
variables
\be
\label{sepvar}
h_{\mu\nu}(x^\lambda, y^a) = \sum_m \Phi_m(y^a)\,\gamma^{(m)}_{\mu\nu}(x^\lambda) \ ,
\ee
where $\gamma_{\mu\nu}^{(m)}(x)$ are the purely 4-dimensional spin-2 fields of mass $m$, 
satisfying
\be
\Box_{(4)} \gamma^{(m)}_{\mu\nu} = m^2\,\gamma^{(m)}_{\mu\nu}
\ee
with $\Box_{(4)} = \eta^{\mu\nu} \partial_{\mu} \partial_{\nu}$. The wave equation for the 
radial profile of the modes in the internal dimension may be written
\be
\label{equaphi}
- \hat{\nabla}_c \left(e^{4A}\,\hat{\nabla}^c \Phi_m \right) = m^2\,e^{2A}\,\Phi_m \ ,
\ee
where $\hat{\nabla}_c$ is the covariant derivative associated to the metric in the 
internal space, $\hat{g}_{ab}$ in (\ref{back}). 

The spectrum and profile of the modes are mainly determined by the region where the warp factor 
$e^{2A}$ varies exponentially. We will first consider the KK modes in the local isometric 
background (\ref{isoback}). When isometries are broken according to (\ref{perback}), the metric 
perturbations are suppressed, $\epsilon(y) \sim e^{-\alpha y/ R} \ll 1$,  in the main part 
of the throat. We will solve the KK mode equation around this non-isometric background at linear order 
in $\epsilon$, by using the standard techniques of quantum mechanics perturbation theory 
(sub-section \ref{isobre}).

\subsection{KK Modes of Isometric Throat}
\label{isothro}

To study the spin-2 KK modes around the local geometry (\ref{isoback}), we may use the 
results of the Appendix with $e^{2A(y^c)} = e^{-2y/R}$ and 
$\hat{g}_{ab} dy^a dy^b = dy^2 + R^2\,f_{ij}\,d\theta^i d\theta^j$. 
In this geometry, the mode equation (\ref{equaphi}) reads
\be
\label{isophi}
- \frac{d}{dy} \left[e^{-4y/R}\,\frac{d\Phi_m}{dy}\right] - 
\frac{e^{-4y/R}}{R^2}\,\Delta_{f} \Phi_m = m^2\,e^{-2y/R}\,\Phi_m \ ,
\ee
where $\Delta_{f}$ denotes the Laplace operator associated with the metric 
$f_{ij}(\Omega_5)$ of the angular coordinates. 

The background symmetries lead to the separation of variables
\be
\label{sepvar2}
\Phi_m(y^c) = \psi_{nL}(y)\,Q_L^M(\Omega_5)  
\ee
with 
\be
\label{Delta}
\Delta_{f} Q_L^M = - F^2(L)\,Q_L^M \ ,
\ee
where $L$ denotes collectively the quantum numbers involved in the eigenvalues $F^2(L)$ 
of $\Delta_{f}$, and $M$ the other ones. For instance, if $f_{ij}$ is the metric of the 
2-sphere, $Q_L^M$ are the usual spherical harmonics $Y_L^M$ and $F^2 = L(L+1)$, where $L$ 
are positive integers and $M$ are integers with $|M| \leq L$. For a Klebanov-Strassler throat, 
$Q_L^M$ would be the harmonics associated to the angular dimensions of the deformed conifold. 

Substituting (\ref{sepvar2}) in Eq. (\ref{isophi}) and using (\ref{Delta}) gives the radial 
wave equation
\be
\label{isopsi}
- \frac{d}{dy} \left[e^{-4y/R}\,\frac{d\psi_{nL}}{dy}\right] + 
\frac{e^{-4y/R}}{R^2}\,F^2(L)\,\psi_{nL} = m_{nL}^2\,e^{-2y/R}\,\psi_{nL} \ ,
\ee
whose eigenfunctions $\psi_{nL}$ and eigenvalues $m_{nL}^2$ will depend on 
the quantum number(s) $L$ (through $F^2(L)$) and on a radial quantum number $n$. For the 
massive modes ($m^2 \neq 0$), the solution for the radial wave-function is
\be
\label{psi}
\psi_{nL}(y) = N_{nL}\,e^{2y/R}\,\left[J_\nu(m_{nL}Re^{y/R}) - B_{nL}\,Y_\nu(m_{nL}Re^{y/R})\right] \ ,
\;\;\; \, \, \,  \;\;\; \nu = \sqrt{4+F^2(L)} \ ,
\ee
where $J_{\nu}$ and $Y_{\nu}$ denote the Bessel functions of order $\nu$ of the first and 
second kind. The constant $B_{nL}$ and the mass spectrum $m_{nL}$ are determined by the 
boundary conditions, while the constant $N_{nL}$ is fixed by the normalization condition. 
In general, KK modes are strongly localized at the bottom of the throat, $y = y_t$, 
and their masses are of the order $m \sim \frac{e^{-y_t/R}}{R}$ \footnote{ Some
KK modes of mass $m \sim V_6^{-1/6}$ may be localized instead in the bulk of the CY, see e.g. \cite{Lykken:2000wz}. 
We will not consider such modes here because for sufficient warping they are much heavier: 
$V_6^{-1/6} \gg e^{-y_t / R} / R$.}.

Boundary conditions may be obtained by matching the solution (\ref{psi}) to the solution 
for the modes at the base  (at $y=0$) and at the tip of the throat (at $y=y_t$).
The effect of the CY on the boundary condition is difficult to model since we don't know the explicit 
metric of the gluing region. Here we will focus on the modes in the interval $0 \leq y \leq y_t$ and for 
simplicity impose Neumann boundary conditions at both ends. To the order of magnitude we will be eventually 
interested in, we don't expect the results to be very sensitive to the particular form of the boundary 
conditions. The conditions $\psi_{nL}'(0) =\psi_{nL}'(y_t) = 0$ translate into two conditions for the $B_{nL}$
\begin{eqnarray}
\label{bc}
B_{nL} \,& =& \, 
\frac{(\nu-2)\,J_\nu(m_{nL}R) - m_{nL}R\,J_{\nu-1}(m_{nL}R)}{(\nu-2)\,Y_\nu(m_{nL}R)-m_{nL}R\,Y_{\nu-1}(mR)} \nonumber \\ 
& = & \frac{(\nu-2)\,J_\nu(m_{nL}R e^{y_t/R}) - m_{nL}R\,e^{y_t/R}\,J_{\nu-1}(m_{nL}R e^{y_t/R})}
{(\nu-2)\,Y_\nu(m_{nL}R e^{y_t/R}) - m_{nL}R\,e^{y_t/R}\,Y_{\nu-1}(m_{nL}R e^{y_t/R})} \ .
\end{eqnarray}

In the following, we will be interested in the lightest among massive KK modes, for which $mR \ll 1 \,$. 
In this case, the first equation above implies $B_{nL} << 1$. More precisely,  $B_{nL} \propto (mR)^2$ for 
$\nu = 2$ and $B_{nL} \propto (mR)^{2\nu}$ for $\nu > 2$. The second equation in (\ref{bc}) 
then reduces approximately to
\be
\label{quant}
(2-\nu)\,J_\nu(\xi_{nL}) + \xi_{nL}\,J_{\nu - 1}(\xi_{nL}) = 0 \ ,
\ee
which determines the spectrum of the KK eignemasses. We have defined $\xi_{nL}$ according to
\be
\label{defxi}
m_{nL} = \frac{\xi_{nL}}{R}\,e^{-y_t/R} \ ,
\ee
where $n = 1, 2, 3, ....$ counts the successive roots of Eq.~(\ref{quant}) for a given $\nu$. 
The mass of the KK modes increases with $n$ and with $\nu$, i.e. with $F^2(L)$ in (\ref{Delta}). 
For $\xi_{nL} \gg 1$ we find an analytical approximated solution (obtained by expanding the 
Bessel functions for large argument)
\be
m_{nL} \simeq \pi \left( \frac{\nu}{2} + n + \frac{1}{4} \right) \, \frac{{\rm e}^{-y_t / R}}{R} \ .
\ee

When the throat is approximated by a patch of $AdS_5 \times T^{1,1}$, we can use the 
eigenvalues of the Laplacian for the 
scalar harmonics (\ref{Delta}) on $T^{1,1}$, see e.g.~\cite{Gubser:1998vd, Jatkar:1999zk, Baumann:2006th}.
We have 
\be
F^2(L) = 6\,\left(l_1 (l_1 + 1) + l_2 (l_2 + 1) - \frac{r^2}{8}\right) \ ,
\ee
where $L = (l_1, l_2, r)$ are the $SU(2)_1 \times SU(2)_2 \times U(1)$ principal quantum numbers, 
with $l_1$ and $l_2$ both integers or both half-integers, and with $r/2 \in \{-l_1, ..., l_1 \}$ and 
$r/2 \in \{-l_2, ..., l_2 \}$.
In this case we can calculate the spin 2 mass spectrum of the angular KK modes.
For illustration, we list in Table I the lightest massive states when the throat is approximated by a patch of 
$AdS_5 \times T^{1,1}$ with Neumann boundary conditions for the radial wave functions.
They are tabulated according to their radial quantum number $n$ and their angular quantum numbers $l_1, l_2, r$.

\begin{table}[htb]
\label{SpecT11}
\begin{center}
\begin{tabular}{|c|c|c|c|c|c|}
\hline\hline
\noalign{\smallskip}
$\;n\;$ & $\;l_1\;$ & $\;l_2\;$ & $\;|r|\;$ & $\;F^2(L)\;$ & $\;\xi_{nL}\;$ \\
\hline\hline
\noalign{\smallskip}
$\;1\;$ & $\;0\;$ & $\;0\;$ & $\;0\;$ & $\;0\;$ & $\;3.83\;$ \\[1mm]
$\;1\;$ & $\;1/2\;$ & $\;1/2\;$ & $\;1\;$ & $\;33/4\;$ & $\;5.45\;$ \\[1mm]
$\;1\;$ & $\;1\;$ & $\;0\;$ & $\;0\;$ & $\;12\;$ & $\;5.98\;$ \\[1mm]
$\;1\;$ & $\;0\;$ & $\;1\;$ & $\;0\;$ & $\;12\;$ & $\;5.98\;$ \\[1mm]
$\;2\;$ & $\;0\;$ & $\;0\;$ & $\;0\;$ & $\;0\;$ & $\;7.02\;$ \\[1mm]
$\;1\;$ & $\;1\;$ & $\;1\;$ & $\;2\;$ & $\;21\;$ & $\;7.04\;$ \\[1mm]
$\;1\;$ & $\;1\;$ & $\;1\;$ & $\;0\;$ & $\;24\;$ & $\;7.35\;$ \\[1mm]
$\;1\;$ & $\;3/2\;$ & $\;1/2\;$ & $\;1\;$ & $\;105/4\;$ & $\;7.57\;$ \\[1mm]
$\;1\;$ & $\;1/2\;$ & $\;3/2\;$ & $\;1\;$ & $\;105/4\;$ & $\;7.57\;$ \\[1mm]
$\;1\;$ & $\;3/2\;$ & $\;3/2\;$ & $\;3\;$ & $\;153/4\;$ & $\;8.63\;$ \\[1mm]
$\;2\;$ & $\;1/2\;$ & $\;1/2\;$ & $\;1\;$ & $\;33/4\;$ & $\;8.92\;$ \\[1mm]
\hline\hline
\end{tabular}
\caption{Lightest massive spin 2 KK modes of the throat approximated by a patch of $AdS_5 \times T^{1,1}$ 
with Neumann boundary conditions for the radial wave functions.}
\end{center}
\end{table}

Now we compute the normalisation constant $N_{nL}$. The normalisation condition for 
each $\gamma^{(m)}_{\mu\nu}(x)$ in (\ref{sepvar}) to represent a canonically normalized 
spin-2 field from the 4-dimensional point of view is given in Eq.~(\ref{orthonorm}). 
We normalize the wavefunctions in the angular directions according to
\be
\label{orthQ}
\int d^5\Omega_5\,Q_L^M(\Omega_5)\,Q_{L'}^{M'}(\Omega_5) \, = \, \delta_{LL'}\,\delta_{MM'} \ ,
\ee
where $d\Omega_5 = d\theta^1 \, ... \,  d\theta^5 \, \sqrt{f}$ is the element of solid angle. 
Eq.~(\ref{orthonorm}) (here for $D = 10$) then gives 
\be
\label{orthpsi}
\frac{M_{10}^8 R^5}{4}\;\int_0^{y_t} e^{-2y/R}\,\psi_{nL}(y)\,\psi_{n'L}(y)\,dy \, 
= \, \delta_{n n'}
\ee
for the radial profile of the modes,
where $M^8_{10}=\frac{2}{(2\pi)^7 g_s^2 \alpha'^4}$.
In the range of masses we are interested in
(${\rm e}^{-y_t / R} < m R \ll 1$), this integral is dominated by the 
upper limit, where we can neglect the term in $B_{nL}\,Y_{\nu}$. This gives
\be
\label{NmL}
N_{nL} \simeq \frac{2 \sqrt{2}}{M_{10}^{4}\,R^3} \, 
\frac{e^{-y_t/R}}{J_\nu \left( \xi_{nL} \right)}\,
\left(1-\frac{F^2}{\xi_{nL}^2}\right)^{-1/2} \ ,
\ee
see e.g. \cite{abraste}. It is more convenient to rewrite this in terms of the 
$4$-dimensional Planck mass, by using Eq.~(\ref{mp}):
\be
\label{NmL2}
N_{nL} \simeq \frac{2 \sqrt{2}}{\Mp} \left( \frac{V_6}{R^6} \right)^{1/2} \, 
\frac{e^{-y_t/R}}{J_\nu \left( \xi_{nL} \right)} \ ,
\ee
where we have also neglected the last term in parenthesis 
(which is approximately equal to one). 

For the modes with angular momentum, the second term in (\ref{psi}) is comparable to the first one 
at $y = 0$ and becomes quickly negligible as $y$ increases. Their normalized wave function is thus 
well approximated by 
\be
\label{psiy}
\psi(y) \simeq \frac{2 \sqrt{2}}{\Mp} \left( \frac{V_6}{R^6} \right)^{1/2} \, 
\frac{J_\nu \left(m_{nL} R e^{y/R}\right)}{J_\nu \left( \xi_{nL} \right)} \, 
\frac{e^{2y/R}}{e^{y_t/R}}\ .
\ee

Finally, Eq.~(\ref{isopsi}) has a solution with $m_{nL}^2 = 0$ if $F^2(L) = 0$. This mode, 
with zero mass and no angular momentum, is the usual 4-D graviton. Its wave function is 
constant in the extra dimensions and normalized to
\be
\label{zeromode}
\Phi_0 = \psi_{00}\,Q_0^0 = \frac{2}{\Mp} \, ,
\ee
see (\ref{graviton}) (for this specific mode, the the leading contribution to the normalisation constant 
comes from the bulk of the CY).

\subsection{KK Modes of the Throat with Isometry Breaking Perturbations}
\label{isobre}

We now study the spin 2 KK modes in the background geometry (\ref{perback}), where the 
isometry breaking perturbations are suppressed by $\epsilon(y) = e^{-\alpha y/R}$ in the throat.
The corresponding mode equation is given by Eq.~(\ref{equaphi}) with 
$e^{2A(y^c)} = e^{-2y/R}\,\left[1 + \epsilon(y)\,w(\Omega_5)\right]$ and 
$\hat{g}_{ab} dy^a dy^b = dy^2 + R^2\,\left[f_{ij}(\Omega_5) + \epsilon(y)\,\delta f_{ij}(\Omega_5)\right]$. 
The function $\epsilon(y)$ is much smaller than one in the throat itself and we
can treat it as a perturbation.
Up to first order with respect to $\epsilon$, the mode equation may be written as
\be
\label{schropert}
\left[H_0 + V\right]\,\Phi_{m} = m^2\,e^{-2y/R}\,\Phi_{m} \ ,
\ee
where
\be
H_0 = - \frac{\partial}{\partial y}\left[e^{-4y/R}\,\frac{\partial}{\partial y}\right] 
- e^{-4y/R}\,\frac{\nabla_{(f)}^2}{R^2} 
\ee
is the unperturbed operator involved in the LHS of the mode equation (\ref{isophi}) for the 
isometric background, while
\begin{eqnarray}
\label{Vpert}
V = \epsilon\,w\,H_0 &-& e^{-4y/R}\,\frac{d \epsilon}{dy}\,\left(2 w + \frac{1}{2} f^{kl} \delta f_{kl}\right)\,
\frac{\partial}{\partial y} \nonumber \\
&-& \frac{e^{-4y/R}}{R^2}\,\epsilon\,\left[\partial_i\left(2 w + \frac{1}{2} f^{kl} \delta f_{kl}\right)\,f^{ij}\,\partial_j - 
\frac{1}{\sqrt{f}}\,\partial_i\left(\sqrt{f} \delta f^{ij} \partial_j\right)\right]
\end{eqnarray}
is the perturbation operator encoding the effects of isometry breaking.

To solve for Eq.~(\ref{schropert}), we may follow the standard procedure of perturbation 
theory, by expanding the wave function $\Phi_m$ on the complete set $\psi_{nL}(y)\,Q_L^M(\Omega_5)$ of 
solutions of the unperturbed problem of the previous sub-section, and determine the 
coefficients of the expansion by using the orthonormal conditions (\ref{orthQ}, \ref{orthpsi}).
We then define the matrix elements 
\be
\label{Vmat}
V_{nLn'L'}^{MM'} = \int dy\,d\Omega_5\,\psi_{nL}(y)\,Q^{M}_{L}(\Omega_5)\,
V\,\psi_{n'L'}(y)\,Q^{M'}_{L'}(\Omega_5) 
\ee
of the perturbation operator $V$ in the basis $\{\psi_{nL} Q_L^M \}$. 
The eigenvalues at zeroth order in $\epsilon$, i.e. the KK masses squared 
$m_{nL}^2$ in the isometric background, will typically be degenerate since in 
particular they do not depend on the quantum number(s) $M$ of the harmonics $Q_L^M(\Omega_5)$. 
In this case, the correct wave functions at zeroth order are the linear combinations which 
diagonalize the matrix elements of degenerate states (see e.g.~\cite{LL}): $V_{nLn'L'}^{MM'} = 0$ 
for $m_{nL} = m_{n'L'}$ and $(n,L,M) \neq (n',L',M')$. We will keep the notation $\psi_{nL} Q_L^M$ 
for these correct wave functions at zeroth order. 

The KK modes' masses up to first order in $\epsilon$ are then given by 
\be
m_{nLM}^2 = m_{nL}^2 + \frac{1}{4} \,M_{10}^8 R^5   \,V_{nLnL}^{MM} \ ,
\ee
where $m_{nL}$ are the masses at zeroth order considered in the previous sub-section. In general, 
the breaking of the isometries removes the degeneracy. We will be specially interested in the 
wave functions of the KK modes when the isometries are broken. Up to first order in $\epsilon$, 
they are given by~\footnote{The matrix (\ref{Vmat}) is not symmetric, but one may derive the relation
\be
V_{n'L'nL}^{M'M} - V_{nLn'L'}^{MM'} \, = \, \left( m_{n'L'}^2 - m_{nL}^2 \right)\,\int dy\,d\Omega_5\,
e^{-2y/R}\,\epsilon\,\left(w + \frac{1}{2} f^{kl}\,\delta f_{kl}\right)\,\psi_{nL}\,Q_L^M\,\psi_{n'L'}\,Q_{L'}^{M'} \ .
\ee
From this it can be shown that the wave functions (\ref{phipert}) satisfy the orthonormal 
conditions (\ref{orthonorm}) at first order in $\epsilon$, as they should. This fixes the constant 
in $\mathcal{O}(\epsilon)$ in the first term of (\ref{phipert}), but its expression will not be 
important for what follows.}
\be
\label{phipert}
\Phi_{nL}^M(y, \Omega_5) = \left(1 + \mathcal{O}(\epsilon)\right)\,\psi_{nL}(y)\,Q_L^M(\Omega_5) \, + \, \frac{1}{4}
\sum_{\stackrel{n', L', M'}{m_{n'L'} \neq m_{nL}}} \frac{M_{10}^8 R^5 \,V_{n'L'nL}^{M'M}}{m_{nL}^2 - m_{n'L'}^2}
\,\psi_{n'L'}(y)\,Q_{L'}^{M'}(\Omega_5) + ... \ ,
\ee
where the ellipses stand for similar contributions due to the matrix elements between degenerate states 
(see \cite{LL}).

Of course, there is still a zero mode 4-D graviton, with zero mass and constant wave function, 
in the absence of isometries in the internal space. Indeed for this mode, the corrections to 
the zeroth order mass and wave function vanish: $V^{M0}_{nL00} = 0$ because $\psi_{00} Q_0^0$ is 
constant. 

Consider now the matrix elements $V_{n'L'nL}^{M'M}$ between two massive states. 
The contribution of the first term in (\ref{Vpert}) to (\ref{Vmat}) may be written 
\be
\label{VnLnL}
V_{n'L'nL}^{M'M} = m_{nL}^2 \, \int_0^{y_t} dy\,e^{-2 y/R}\,\epsilon(y)\,\psi_{nL}(y)\,\psi_{n'L'}(y) 
\, \int d\Omega_5\,w(\Omega_5)\,Q_L^M(\Omega_5)\,Q_{L'}^{M'}(\Omega_5)+ ... \ .
\ee 
The other terms in (\ref{Vpert}) lead to similar contributions.
The integrals over the angles determine the transitions allowed (i.e. the non-vanishing 
matrix elements), while the $y$-integral determine their amplitude. 

In the following, we will 
need the amplitude of the matrix element between two light massive states 
($m_{nL} \sim m_{n'L'} \sim e^{-y_t/R} / R \neq 0$) with non-zero angular momentum ($L, L' \neq 0$). 
Using the expression (\ref{psiy}) for the $\psi_{nL}$ and with the change of variables $u = m_{nL}R e^{y_t/R}$, 
the $y$-integral in (\ref{VnLnL}) for a perturbation $\epsilon(y) = e^{-\alpha y / R}$ gives
\ba
\label{integral}
&&\int_0^{y_t} dy\,e^{-(2+\alpha)y_t/R}\,\psi_{nL}(y)\,\psi_{n'L'}(y) = \nonumber \\
 &=& \frac{8}{\Mp^2}\,\frac{V_6}{R^6}
\,\frac{R \, \xi_{nL}^{\alpha-2} \, e^{-\alpha y_t/R}}{J_{\nu}(\xi_{nL}) J_{\nu'}(\xi_{n'L'})} \, 
\int_{\xi_{nL} e^{-y_t/R}}^{\xi_{nL}} du\,
u^{1-\alpha}\,J_{\nu}\left(u\right)\,J_{\nu'}\left(\frac{\xi_{n'L'}}{\xi_{nL}} u\right) 
\approx \frac{e^{-\alpha y_t /R}}{\Mp^2}\,\frac{V_6}{R^5} \ ,
\ea   
where we have used (\ref{defxi}) and $\xi_{nL} \sim \mathcal{O}(1)$ for the 
lightest modes. For $u \rightarrow 0$, $J_{\nu}(u) \propto u^{\nu}$, so the u-integral does not depend on its lower 
limit for $\alpha < 2 + \nu + \nu'$.\footnote{For $\alpha > 2 + \nu + \nu'$, the $u$-integral diverges for 
$u \rightarrow 0$ and is then dominated by its lower limit, leading to
\be
\label{lowlim}
\int_0^{y_t} dy\,e^{-(2+\alpha) y_t/R}\,\psi_{nL}(y)\,\psi_{n'L'}(y) \approx 
\frac{e^{-(2+\nu+\nu') y_t /R}}{\Mp^2}\,\frac{V_6}{R^5} \; . 
\ee
This case may be covered by replacing $\alpha$ with $\mathrm{Min}(\alpha\, , \, 2 + \nu + \nu')$ 
in Eq. (\ref{phinLM}).} The integral is then mainly accumulated around its upper limit, 
leading to the last equality in (\ref{integral}).

Therefore, for the non-vanishing matrix elements $V_{n'L'nL}^{M'M}$ between two light massive modes with 
angular momentum, the component of the first order wave function $\Phi_{nL}^M$ in (\ref{phipert})
along the zeroth order wave function $\psi_{n'L'} Q_{L'}^{M'}$ is given by
\ba
\label{phinLM}
\Phi_{nL}^M(y, \Omega_5) &=& \frac{1}{4} \frac{M_{10}^8 R^5 \,V_{n'L'nL}^{M'M}}{m_{nL}^2 - m_{n'L'}^2}
\,\psi_{n'L'}(y)\,Q_{L'}^{M'}(\Omega_5)+ ... \nonumber \\ 
&\approx& \frac{e^{-(1+\alpha) y_t/R}}{\Mp}\,\left(\frac{V_6}{R^6}\right)^{1/2}\,
e^{2y/R}\,J_{\nu'}\left(m_{n'L'} R e^{y/R}\right)\,Q_{L'}^{M'}(\Omega_5) +... \ ,
\ea
where we have used (\ref{mp}) and (\ref{defxi}), and keep the contribution of one of the terms in the sum (\ref{phipert}). 
Note that the first order correction to the wave function is smaller than the wave function at zeroth order 
by a factor of $\epsilon(y_t)$ at any position in the radial coordinate $y$.

\section{Interactions of KK Modes} \label{sec:interactions}

In this Section, we will study interactions of the KK modes.
We shall include 3-legs and 4-legs self-interactions of the KK modes.  
Most important for cosmology will be interactions of KK modes with the Standard Model particles.
This will depend on the SM phenomenology in the string theory setting. Following \cite{lev}, we will consider
a simple model of KK modes interacting with the SM located on a 3-brane around the tip of the same throat.
Notice that this model differs from what was considered in \cite{Berndsen:2007my}. 
The cases where the Standard Model degrees of freedom are located in the bulk 
of the CY or in another throat are further constrained and will be discussed in the last, 
conclusion Section. 

The decay times of angular KK modes due to isometry breaking are much longer than the time scales for 
their annihilations. It is convenient to call them as the  ``late'' decays due to isometry breaking versus 
``early'' decays and annihilations which are taking place without isometry breaking.
Therefore, we consider first the ``early''  channels of KK interactions, for which 
the isometry breaking of the throat can be neglected. At this level we will deal with 
KK modes with conserved angular momenta (of inner isometries).  

We first compute the coupling constants for KK modes self-interactions in sub-section \ref{early}.
In sub-section \ref{annihil}, we then compute the coupling constants for the KK modes interactions 
with the SM at the brane, again, with no isometry breaking from the CY. In both cases, we pay 
special attention to the selection rules to be satisfied.  
This will allow us to identify the long-living angular KK modes, which could pose cosmological problems. 
We then study in sub-section \ref{late} the ``late'' decays of the long-lived KK modes induced by the 
isometry breaking perturbations.

\subsection{Self-interactions of KK Modes without Isometry Breaking Perturbations} 
\label{early}

For the 
decays and annihilations of KK modes into themselves, 
the effective four-dimensional Lagrangian  and  coupling constants 
are obtained by substituting the perturbations (\ref{spin2}) into the bulk action, expanding to third 
and fourth order in $h_{\mu\nu}$, and integrating over the internal coordinates. 
This is done in details in the Appendix for the general background (\ref{back}), 
up to third order in $h_{\mu\nu}$. 

Here we consider KK modes self-interactions which are already present in the throat
geometry (\ref{isoback}) without isometry breaking, i.e. at zeroth order {\it w.r.t.} 
the isometry breaking parameter $\epsilon(y)$.
 
We start with the decay of a KK mode of mass $m$ and 
quantum numbers $(n, L, M)$ into two lighter KK modes, of masses $m'$ and $m''$, and 
quantum numbers $(n', L', M')$ and $(n'', L'', M'')$. The corresponding terms in the 
effective action have the form 
\ba
\label{3KK}
\frac{M_{10}^{8} R^5}{2} \int e^{-2y/R}\,\psi_{nL}\,\psi_{n'L'}\,\psi_{n''L''}\,dy 
&&\int Q^M_{L}\,Q^{M'}_{L'}\,Q^{M''}_{L''} d\Omega_5 
\int \gamma_\mu^{(m)\nu} \partial_\sigma \gamma^{(m)}_{\nu\rho} \partial^\sigma \gamma_{(m)}^{\mu\rho}\,d^4x \, 
= \nonumber \\
&=& \, \lambda_{3 {\rm KK}} \int \gamma_\mu^{(m)\nu} \partial_\sigma \gamma^{(m)}_{\nu\rho} 
\partial^\sigma \gamma_{(m)}^{\mu\rho}\,d^4x \ ,
\ea
where $d\Omega_5 = d\theta^1 \, ... \,  d\theta^5 \, \sqrt{f}$ is the element of solid angle in the 
angular dimensions $X_5$. The expression above may be obtained by substituting (\ref{sepvar2}) into (\ref{S3}) 
for the background (\ref{isoback}). Here $M_{10}^8$ is the 10-dimensional Planck mass, which is related to 
the 4-dimensional Planck mass $\Mp$ and the total volume of the internal space $V_6$ through Eq. (\ref{mp}). 
With our choice of normalisation for the wave function in the internal space, the $\gamma_{\mu\nu}^{(m)}$ are 
canonically normalized spin 2 fields, and $\lambda_{3 {\rm KK}}$ is the corresponding coupling constant.  
Eq. (\ref{3KK}) allows us to calculate $\lambda_{3 {\rm KK}}$.

Let us consider the selection rules that follows from Eq.~(\ref{3KK}).
The conservation of the 4-dimensional energy and momentum in the center of mass frame reads as
\be
\label{Econs}
m = \sqrt{m'^2 + \mathbf{k}^2} + \sqrt{m''^2 + \mathbf{k}^2} \ ,
\ee
which requires in particular $m > m' + m''$. 

The integral over $d\Omega_5$ imposes angular momentum selection rules.
By contrast, the 
warped background has no translation invariance in the radial direction $y$, so the $y$-integral 
does not impose the conservation of the total KK modes mass: $m$ may be different from $m' + m''$,  
contrary to the case of  extra dimensions with translation invariance. However, if one of the decay product is 
the massless graviton, $m'' = L'' = M'' = 0$, its wave function is constant in the internal 
space and drops out of the integrals over the inner dimensions. With (\ref{orthQ}), the $y$-integral then reduces 
to the orthogonality conditions (\ref{orthpsi}), which implies $m = m'$. 
Furthermore, decay of KK modes is cascading into lighter and lighter KK states. However, if one of the decay 
product is the zero-mode graviton, $m'' = 0$, then $m = m'$ is required, as we just saw above and is shown more 
generally in the Appendix (see Eqs.~(\ref{S3}) and (\ref{S23})). 
Therefore, a massive KK mode ($m \neq 0$) cannot decay into two massless gravitons $m' = m'' = 0$ 
(this also was noticed in \cite{tye2}). It is also obvious that a massive KK mode cannot decay into one massless graviton 
and another massive mode since the condition $m = m'$ leaves no phase space for this process.  
Therefore, a massive KK mode cannot decay by 3-legs interactions into one or two zero-mode graviton(s). 
As shown in the Appendix, this is independent of any isometries in the internal space.
In addition, suppose  $m_0$ is the mass of the lightest massive KK mode ($m_0 \neq 0$).
Evidently, only the modes with $m > 2 m_0$ can decay into two other massive modes. 
We thus conclude that the modes with $m \leq 2 m_0$ cannot decay by 3-legs interactions into any other spin 2 modes 
(massive or massless). 

Finally, in the presence of isometries, the conservation of the corresponding angular momentum imposes further 
selection rules. In particular, a mode with non-zero angular momentum ($L \neq 0$) cannot decay into modes with 
non-zero angular momentum only ($L' = L'' = 0$), which implies that the 
lightest modes with non-zero angular momentum cannot decay into any other KK mode.
  
For the allowed interactions between $3$ massive KK modes, the effective 4-dimensional coupling 
constant is obtained by substituting Eq.~(\ref{psiy}) into the {\it l.h.s} of Eq.~(\ref{3KK}), and 
performing the integrals over the internal coordinates. The $y$-integral is accumulated mostly  
around $y = y_t$, leading to
\be
\label{lam3KK}
\lambda_{3 {\rm KK}} \, \approx \, \left(\frac{V_6}{R^6}\right)^{1/2}\,\frac{e^{y_t/R}}{\Mp} \ ,
\ee
where we have neglected numerical factors of order unity and used (\ref{mp}) to express $M_{10}^8$ 
in terms of $\Mp$ and $V_6$.  

The coupling constants for KK modes 4-legs interactions are obtained in the same way as we did 
it above  for the 3-legs interactions. Expansion of the bulk action to the fourth order 
{\it w.r.t.}  $h_{\mu\nu}$ gives the effective coupling constant for interactions between 
four KK modes
\be
\lambda_{4 {\rm KK}} \, \approx \, M_{10}^{8} R^5 \, \int_0^{y_t} e^{-2y/R}\,
\psi_{nL}\,\psi_{n'L'}\,\psi_{n''L''}\,\psi_{n'''L'''}\,dy \ .
\ee
For four massive modes, the integral is again dominated by the contribution around its
upper limit, leading to
\be
\label{lam4KK}
\lambda_{4 {\rm KK}} \, \approx \, \left(\frac{V_6}{R^6}\right)\,\frac{e^{2 y_t/R}}{\Mp^2} \ .
\ee
It is worth noticing that this is about the square of (\ref{lam3KK}),  $\lambda_{4 {\rm KK}} \approx \lambda_{3 {\rm KK}}^2$.

\subsection{KK-SM particles Interactions Without Isometry Breaking Perturbations}
\label{annihil}

To obtain the coupling constants for the decays and annihilations of KK modes into SM brane 
degrees of freedom, we consider  4-dimensional fields located on a probe brane at $y^c = y_b^c$ in the geometry (\ref{back}).
One can start with the DBI action for the probe brane and use the decomposition of the DBI action for the
perturbative field fluctuations. For simplicity, for a light scalar four-dimensional brane degree of freedom $H$, we have 
\be
\label{Sb}
\int d^Dx\,\sqrt{-G_{(\mathrm{ind})}}\,G_{(\mathrm{ind})}^{\mu\nu}\,\partial_{\mu}H\,\partial_{\nu}H\;
\delta^{(6)}(y^c-y_b^c)\, = \, 
\int d^4x\,\sqrt{-g(y_b^c)}\,g^{\mu\nu}(y_b^c)\,\partial_{\mu}\hat{H}\,\partial_{\nu}\hat{H} \ , 
\ee 
where $G^{ind}_{\mu\nu} = e^{2A(y_b^c)}\,g_{\mu\nu}(y_b^c)$ is the induced metric on the brane located at 
$y^c = y_b^c$. 
To zeroth order in $h_{\mu\nu}$, the kinetic term is canonical for the normalized scalar field 
$\hat{H} = e^{A(y_b^c)}\,H$, that we have substituted in the {r.h.s.} of Eq.~(\ref{Sb}). The interaction 
term $h^{\mu\nu} \,\partial_{\mu}\hat{H}\,\partial_{\nu}\hat{H}$
 is  then  obtained by substituting $g_{\mu\nu} = \eta_{\mu\nu} + h_{\mu\nu}$, where
$h_{\mu\nu}$ are the spin 2 KK modes (\ref{spin2}). 
 Later on, we will consider the branching 
ratios for the KK modes decay into different Standard Model fields. For the moment we are only interested 
in the overall coupling constant, which can  be estimated for the simplest case  of a light scalar field.

Consider the decay of a KK mode with quantum numbers $(n, L, M)$ in (\ref{sepvar2})
into two lighter $3$-brane scalar degrees of freedom, KK $\to$ b b, following from (\ref{Sb}).
It involves the term
\be
\label{KKb}
Q_L^M(\Omega_b)\,\psi_{nL}(y_b)\,\int d^4x\,
\gamma_{(m)}^{\mu\nu}\,\partial_{\mu}\hat{H}\,\partial_{\nu}\hat{H} 
= \lambda_{{\rm KK}\,{\rm b}\,{\rm b}}\,\int d^4x\,\gamma_{(m)}^{\mu\nu}\,\partial_{\mu}\hat{H}\,\partial_{\nu}\hat{H} \ ,
\ee
where $\Omega_b$ and $y_b$ denote the position of the brane in the angular and radial dimensions respectively.

At first glance, the coupling constant $\lambda_{{\rm KK}\,b\,b}$ somehow 
depends on the value of the angular eigenmode of the KK wave function at the position of the brane.
While the brane will break some, but not all, the isometries of the bulk, the interaction 
(\ref{KKb}) should respect the conservation of angular momentum associated with the remaining isometries.

\begin{figure}[h]
\centerline{
\includegraphics[width=0.6\textwidth]{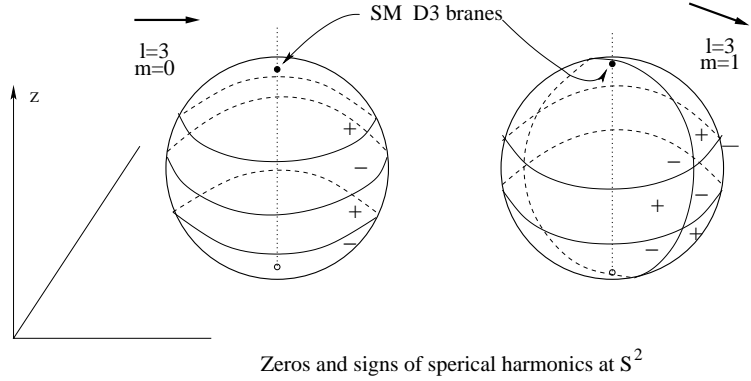}
}
\caption{Zeros and signs on the spherical harmonics on $S^2$. We compare the $M=0$ with the $M\neq0$ case.
 The spherical harmonics with $M \neq 0$ vanish at the two poles. As a consequence, $KK$ 
modes with nonvanishing angular momentum along the directions whose isometries are left
 unbroken by the brane are not directly coupled to brane fields (see the main text for details).}
\label{sign}
\end{figure}

To illustrate that the interaction in the form (\ref{KKb}) indeed conserves quantum numbers associated to 
the isometries unbroken by the brane, we consider an example where the bulk involves an exact 2-sphere $S^2$. 
The setup is shown in Fig.~\ref{sign}. The angular dependence of the KK mode wavefunction on $S^2$ is given by 
the usual spherical harmonics, $Q_L^M \propto Y_L^M(\theta, \phi)$. A $3$-brane breaks the $SO(3)$ isometry of 
the $S^2$ into $SO(2)$. Choosing, for instance, the $z$-axis to point in the direction of the position of the 
brane on the $S^2$ (here the north pole, see Fig. \ref{sign}), $\theta_b = 0$, the coupling constant in (\ref{KKb}) 
is proportional to $\lambda_{{\rm KK}\,b\,b} \propto Y_L^M(\theta_b = 0) \propto P_L^M(0)$ (where $P_L^M$ are the 
associated Legendre polynomials \cite{abraste}). Notice that this vanishes for all $M \neq 0$, see Fig. \ref{sign} 
(where $M$ is the projection of the angular momentum along the $z$-axis). 
This means that only the modes with $M = 0$ can decay. The conservation of the quantum number $M$ associated to 
the isometries unbroken by the brane forbids the modes with $M \neq 0$ to decay. Indeed, for these modes, the angular 
momentum vector has a component along the $z$-axis, and rotations around this axis are left invariant by the brane. 
Analogously, a 3-brane on the $3$-sphere at the tip of a Klebanov-Strassler throat breaks the $SO(4)$ isometry 
into $SO(3)$, and KK modes with non-zero angular momentum associated with the remaining isometries cannot decay.

For the allowed interactions, the coupling constant $\lambda_{{\rm KK}\,{\rm b}\,{\rm b}}$ in (\ref{KKb}) 
is obtained by evaluating the wave function (\ref{psiy}) at the position of the brane, $y=y_b$. For the decay of 
a massive KK mode ($m_{nL} \neq 0$), this gives 
\be
\label{lamKKbb}
\lambda_{{\rm KK}\,{\rm b}\,{\rm b}} \, \approx \, \left(\frac{V_{6}}{R^{6}}\right)^{1/2}\,
\frac{e^{(2 y_b - y_t)/R}}{\Mp}\,\frac{J_\nu(mRe^{y_b/R})}{J_\nu(mRe^{y_t/R})} \ .
\ee
Comparing with Eq. (\ref{lam3KK}), we see that the 3-legs coupling of a KK mode with brane degrees 
of freedom located at the tip of the throat, $y_b = y_t$, is of the same order as the 3-legs coupling 
of the KK modes between themselves. We will denote this coupling constant by $\lambda_3$ in the following
\be
\label{lam3}
\lambda_{3 {\rm KK}} \, \approx \, \lambda_{{\rm KK}\,{\rm b}\,{\rm b}} \, \approx \, 
\left(\frac{V_6}{R^6}\right)^{1/2}\,\frac{e^{y_t/R}}{\Mp} \, \equiv \, \lambda_3 \ .
\ee

Interactions involving two KK modes and two brane degrees of freedom are obtained by expanding 
(\ref{Sb}) to the second order in $h_{\mu\nu}$. The corresponding coupling constant is obtained in the 
same way as in (\ref{KKb}). Again, it vanishes for KK modes with non-zero angular momentum associated 
with the isometries unbroken by the brane. For two other, massive KK modes, it is given by
\be
\label{lam2KKbb}
\lambda_{{\rm KK}\,{\rm KK}\,{\rm b}\,{\rm b}} \, \approx \, \psi_{nL}(y_b) \, \psi_{n'L'}(y_b) \, \approx \, 
\left(\frac{V_{6}}{R^{6}}\right)\,\frac{e^{2 (2 y_b - y_t)/R}}{\Mp^2}\,
\frac{J_\nu(mRe^{y_b/R})\,J_{\nu'}(m'Re^{y_b/R})}{J_\nu(\xi_{nL})\,J_{\nu'}(\xi_{n'L'})} \ ,
\ee
which again reduces to (\ref{lam4KK}) when the brane degrees of freedom are located at the tip 
of the throat, $y_b = y_t$. We will denote this coupling constant by $\lambda_4$, and it is related 
to $\lambda_3$ in (\ref{lam3}) according to
\be
\label{lam4}
\lambda_{4 {\rm KK}} \, \approx \, \lambda_{{\rm KK}\,{\rm KK}\,{\rm b}\,{\rm b}} 
\, \approx \, \lambda_3^2 \, \equiv \, \lambda_4 \ .
\ee

\subsection{Late Decays of KK Modes due to Isometry Breaking Perturbations}
\label{late}

\begin{figure}[h]
\centerline{
\includegraphics[width=0.6\textwidth]{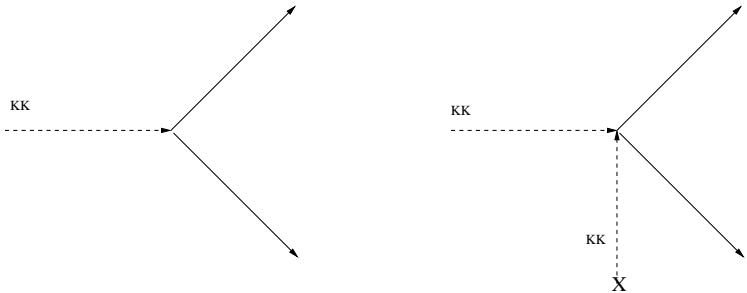}
}
\caption{Left panel: for the isometric throat, the decay of angular KK modes localized around the tip of the throat
is forbidden by higher-dimensional angular momentum conservation.
Right panel: The isometry breaking leads to a new interaction, which corresponds to the decay of an angular KK mode 
into SM particles mediated by a background isometry breaking KK mode localized in the compact CY.}
\label{diag}
\end{figure}

In the isometric throat, the KK modes with non-zero angular momentum associated with the isometries 
unbroken by the brane (the modes with $M \neq 0$ in the $S^2$ example) cannot decay into the brane degrees 
of freedom.
Among these, the lightest ones cannot decay either into any other KK mode. They are therefore 
stable if the isometries are not broken, and become long-lived when the isometries are slightly broken. 
This is illustrated by the diagrams of Fig.~\ref{diag}.

We now study how these long lived modes may decay in the background (\ref{perback}), where the isometry 
breaking perturbations from the CY have radial profile $\epsilon(y) \ll 1$ in the throat. We proceed 
as in the previous sub-sections, but now we consider the effective action at first order in $\epsilon$. 
The corresponding wave functions for the KK modes were calculated in sub-section \ref{isobre}. 

Consider first the interactions between 3 KK modes described by (\ref{S3}). The corresponding 
4-dimensional coupling constant at first order in $\epsilon$ includes terms involving the first 
order correction to $e^{2A}$ and $\sqrt{\hat{g}}$, and 3 zeroth order KK modes wave functions
\be
\label{3KKe}
\lambda_{3\,{\rm KK}}^{(\epsilon)} =\int dy\,e^{-2 y/R}\,\epsilon(y)\,\psi_{nL}\,\psi_{n'L'}\,\psi_{n''L''}\,
\int d\Omega_5\, \left(w + \frac{1}{2} f^{kl} \delta f_{kl}\right)\,Q_{L}^{M}\,Q_{L'}^{M'}\,Q_{L''}^{M''} + ... \ .
\ee
In addition, there are terms involving the unperturbed background metric and the first order $\epsilon$ correction to
 the wave-functions, not shown in (\ref{3KKe}). The complete expression is given by Eqs.~(\ref{S3}) and (\ref{S23}).

In  order to evaluate $\lambda_{3\,{\rm KK}}^{(\epsilon)}$ it is sufficient  to consider the first term in (\ref{3KKe}).
There is, however, an exception. While the first term in (\ref{3KKe}) formally does not forbid to have massless $m=0$
KK mode in that three-legs interaction, the total amplitude in this case
is zero. The reason is that the trilinear combination of the full KK wave functions in  $\lambda_{3\,{\rm KK}}^{(\epsilon)}$,
when one of them is the constant zero-mode wave function, is effectively reduced to a
bilinear combination. This bilinear combination is constrained by the orthonormality conditions of the full wave functions, 
which forbids the three-legs interactions involving one or two zero-mode(s). See the appendix for details. 

In general, the isometry breaking perturbations are decomposed into harmonics $w^{(\mathcal{L})}(\Omega_5)$ and 
$f_{ij}^{(\mathcal{L})}(\Omega_5)$ with different quantum numbers $\mathcal{L}$, each with its own profile in the radial 
direction, see section \ref{isbr} and Eq. (\ref{series}). Consider the selection rules resulting from a given harmonic 
$f_{ij}^{(\mathcal{L})}$ with radial profile $\epsilon(y) = 
e^{-\alpha_\mathcal{L} y/R}$ in (\ref{3KKe}) \footnote{The 
leading perturbation $\delta f_{ij}^{(\mathcal{L})}$ (the one with the lowest $\alpha_\mathcal{L}$) on $AdS_5 \times T^{1,1}$ 
identified in \cite{Aharony:2005ez} corresponds to a traceless tensor mode on $T^{1,1}$, for which 
$f^{kl} \delta f_{kl}^{(\mathcal{L})}$ vanishes. In this case, the contribution of the leading perturbation to 
$\lambda_{3\,{\rm KK}}^{(\epsilon)}$ comes from the perturbed wave-function (through the last term in (\ref{Vpert})) or from 
the perturbation of the warp factor, leading to similar results. Here we consider the contribution of another harmonic 
for illustration and keep $\alpha_\mathcal{L}$ arbitrary.}. The function $f^{kl} \delta f_{kl}^{(\mathcal{L})}$ in the integral 
over the angles breaks the conservation of angular momentum, which may facilitate the decay cascade of the angular KK modes. 
However, for the lightest of them, the relevant process is their decay into modes with zero angular momentum ($L' = L'' = 0$), 
which requires $\int d\Omega\,f^{kl} \delta f_{kl}^{(\mathcal{L})} Q_L^{M} \neq 0$. The perturbation 
$f^{kl} \delta f_{kl}^{(\mathcal{L})}$ having given principal quantum numbers $\mathcal{L}$, 
it allows only the modes with the same angular momentum, $L = \mathcal{L}$, to decay into two modes with 
zero angular momentum. The lightest modes with angular momentum would then decay through the perturbations  
carrying their own quantum numbers. These perturbations may be exponentially suppressed compared to the leading 
one, resulting in exponentially longer lifetimes. Furthermore, as in sub-section \ref{early}, the decay 
of the modes with $m \leq 2\,m_0$ is still kinematically forbidden, since this does not depend on the isometries 
of the internal space (see the Appendix). As illustrated in Table 1, there may be several of these modes. 
Because of these restrictive selection rules, the leading decay channels for the long-lived modes will be into 
brane degrees of freedom.

For the decay channels which are made possible due to a given isometry breaking perturbation with radial profile 
$e^{-\alpha y /R}$ in the throat,  the coupling constant $\lambda_{3\,{\rm KK}}$ is calculated by evaluating 
the $y$-integral in (\ref{3KKe}) as we did in (\ref{integral}). For 3 light massive KK modes, this gives
\be
\label{lamKKe}
\lambda_{3\,{\rm KK}}^{(\epsilon)} \; \approx \; \left(\frac{V_{6}}{R^{6}}\right)^{1/2}\,
\frac{e^{(1-\alpha) y_t/R}}{\Mp} \ .
\ee

We consider now the decay of a KK mode into 2 brane degrees of freedom at first order in 
$\epsilon$. We are interested in the long-lived KK modes, i.e. the lightest modes with non 
zero angular momentum associated to the isometries unbroken by the brane. As in (\ref{KKb}), 
the corresponding coupling constant is just given by the wave function 
evaluated at the position of the brane in the internal space. This vanishes at zeroth order 
for the modes we are interested in, so we have to consider the first order corrections to 
the wave functions $\Phi_{nL}^M$ given in (\ref{phipert}).

The decay occurs through the combined effect of isometry breaking due to the CY and isometry 
breaking due to the $D3$ (or $\bar D3$) brane. The resulting selection rules are less restrictive but 
more involved than for the decay into two KK modes with zero angular momentum. As an illustration, we return to 
the example of a 3-brane located at $\theta_b = 0$ on a 2-sphere ($Q_{L}^{M} = Y_{L}^{M}$). The modes which could 
not decay into brane degrees of freedom in the isometric case were the ones with non-vanishing quantum number $M$. 
In the series (\ref{phipert}), all the terms involving $Y_{L'}^{M'}$ with $M' \neq 0$ still vanish at the 
position of the brane. The 4-dimensional coupling constant is thus given by 
\ba
\label{lambranee}
\lambda_{{\rm KK}\,{\rm b}\,{\rm b}}^{(\epsilon)} \, &= & \, \Phi_{nL}^{M \neq 0}(y_b, \theta_b \, = 0) \, = \nonumber \\
&=& \frac{1}{4} \sum_{\stackrel{n',L'}{m_{n'L'} \neq m_{nL}}} \frac{M_{10}^8 R^5}{m_{nL}^2 - m_{n'L'}^2}\,V_{n'L'nL}^{0 M}
\,\psi_{n'L'}(y_b)\,Y_{L'}^{0}(\theta_b = 0) \, + \, ... \ .
\ea

This coupling constant is nonvanishing if the wave function $\Phi_{nL}^{M \neq 0}$ acquires a non-zero component 
along some harmonic $Y_{L'}^0$ due to the perturbation, i.e. if there exists some $L'$ such that 
$V_{n'L'nL}^{0M} \neq 0$ for $M \neq 0$. Using (\ref{VnLnL}), the contribution of the first term in (\ref{Vpert}) 
to this condition may be written
\be
\label{VnLnL0M}
\int d\Omega_2\,w\,Y_L^M\,Y_{L'}^0 \, \neq \, 0    
\;\;\;\; \mbox{ for } \;\;\;\; M \neq 0 \ .
\ee
Consider the contribution of the leading perturbation (the one which is the less suppressed at the tip of the throat) 
$w^{(\mathcal{L})}$ with given principal quantum numbers $\mathcal{L}$. In our $2$-sphere example, we would have 
$w^{(\mathcal{L})} = \sum_{\mathcal{M} = - \mathcal{L}}^{\mathcal{L}} c_{\mathcal{M}}\,Y_{\mathcal{L}}^{\mathcal{M}}$. 
In this case, the $\Omega_2$-integral in (\ref{VnLnL0M}) is nonvanishing if there exists an $L'$ such that 
$|\mathcal{L} - L| \leq L' \leq \mathcal{L} + L$ with $\mathcal{L} + L' + L$ even, 
and if there exists an $\mathcal{M}$ such that $\mathcal{M} = M$. 
The first condition can always be satisfied, while the second one requires $L \leq \mathcal{L}$. Thus in this case 
the KK modes can decay into SM (brane) degrees of freedom through the leading perturbation if their angular momentum is smaller 
than  the one of the perturbation. On the other hand, the KK modes with greater angular momentum can cascade into KK modes with lower 
angular momentum already at zeroth order in the isometry breaking, if kinematically possible. Wether all the long-lived 
modes can decay this way is clearly very sensitive to the details of the actual KK spectrum. 
The remaining long-lived modes, if any, have to decay through a sub-leading isometry 
breaking perturbation, or at higher order in perturbation theory, resulting in an exponentially longer decay time. 

We note also that, at least if one considers an $\bar D3$-brane, an isometry breaking perturbation $w$ of the warp factor 
acts as a potential for the angular moduli of the brane \cite{Aharony:2005ez}, and therefore may determine its 
position in the internal space. The isometries left invariant by the brane then depends on the perturbation considered, and in 
particular they may correspond to isometries left invariant also by the perturbation. Suppose for instance that the leading 
isometry breaking perturbation has principal quantum number $l=1$ on a $2$-sphere, 
$w = \sum_{\mathcal{M} = - 1}^{1} c_{\mathcal{M}}\,Y_{1}^{\mathcal{M}}$. It is invariant under rotations around a certain axis. 
The $\bar D3$-brane may be driven to the minimum of the potential $w$, which in the present case (i.e. for $l=1$) is 
located on the same axis. Then both the brane and the perturbation leave the system invariant under rotations around this 
axis. In such a case, the long-lived modes have again to decay through another isometry breaking perturbation.

If the throat is approximated by a patch of $AdS_5 \times T^{1,1}$, then the leading isometry breaking perturbation 
is a tensor mode $\delta f_{ij}$ on $T^{1,1}$ with $\alpha \simeq 1.29$, and potentially dangerous spin 2 KK relics are listed 
in rows 2 to 4 and 6 to 11 of Table I. Which of these are long-lived depends on the way the SM brane degrees of freedom are embedded 
into $T^{1,1}$. The modes with $m < 2 m_0$ (i.e. those with $\xi_{nL} < 2 \times 3.83$ in Table I) cannot decay 
directly into 
two other spin 2 KK modes. Their decay into SM brane degrees of freedom through the leading isometry breaking perturbation may be 
allowed by the last term in Eq. (\ref{Vpert}). Otherwise, they have to decay through subleading isometry breaking perturbations. 
For instance, all the modes of Table I can decay into brane degrees of freedom through the perturbations $f^{kl} \delta f_{kl}$ 
(scalar modes on $T^{1,1}$) listed in Table II of \cite{Aharony:2005ez}, which preserve Lorentz invariance and supersymmetry, 
because these perturbations carry the same quantum numbers as the KK modes in question. In this case, $V_{n'0nL}^{0M} \neq 0$ 
so that the perturbed wave-function of the modes has a non-trivial component along $Q_0^0 = \mathrm{constant}$, which never 
vanishes at the position of the brane.

The coupling constant (\ref{lambranee}) for the decay of a massive KK mode into two lighter brane degrees of 
freedom through an isometry breaking perturbation with radial profile $e^{-\alpha y / R}$ may be 
calculated with (\ref{phinLM}) (see also the comment below Eq. (\ref{lowlim}))
\be
\label{lamKKbbe}
\lambda^{(\epsilon)}_{{\rm KK}\,{\rm b}\,{\rm b}} \; \sim \; 
\left(\frac{V_6}{R^6}\right)^{1/2}\;\frac{e^{-(\alpha + 1) y_t/R}}{\Mp}\,e^{2y_b/R}\,J_{\nu'}(m_{n'L'} R e^{y_b/R}) \ .
\ee
In principle, we should sum over all the terms $(n', L', M')$ with non-vanishing matrix element in (\ref{lambranee}), 
but the term with mass $m_{n'L'}$ closer to the mass $m_{nL}$ of the decaying KK mode dominate. Again, (\ref{lamKKbbe}) 
reduces to (\ref{lamKKe}) when the brane is at the tip of the throat, $y_b = y_t$.

To summarize, we argue   that the leading decay channels for the long-lived modes are the ones into SM brane degrees of freedom.
However, the decay of different KK modes may have to be induced by different isometry breaking perturbations of the throat.  
Different isometry breaking perturbations, in turn, lead to different values of $\alpha$ in (\ref{lamKKbbe}).
Note also that throat models with different isometries than the KS throat would lead to different values of $\alpha$. 
For these reasons in the following we will keep $\alpha$ as a free parameter. The value $\alpha \simeq 1.29$ 
for the leading perturbation considered in \cite{Aharony:2005ez} and related to scalar metric perturbations
(shown as the vertical, ``background'' leg at the right panel of Fig~\ref{diag}) will serve as a reference value. 
We will denote by $\lambda_3^{(\epsilon)}$ the coupling constant (\ref{lamKKbbe}) when the brane is at the tip 
of the throat ($y_b = y_t$)
\be
\label{lameps}
\lambda_3^{(\epsilon)} \; \equiv \; \left(\frac{V_{6}}{R^{6}}\right)^{1/2}\,
\frac{e^{(1-\alpha) y_t/R}}{\Mp} \ .
\ee

This is the main formula that we will use to estimate astrophysical effects of the KK modes decay.
Remarkably, the  coupling (\ref{lameps}) can be significantly weaker than the gravitational coupling $1 / \Mp$,
which makes its astrophysical applications much more interesting.
Indeed, while the factor  $\left(\frac{V_{6}}{R^{6}}\right)^{1/2}$ is greater than one, the exponential factor
$e^{(1-\alpha) y_t/R}$  can be exponentially smaller than one. Already for a throat with warp factor $10^{-3}$
and the modes with $\alpha=1.29$, this factor is about $0.1$. Increase of $\alpha$ further decreases this factor, 
for example, for $\alpha=2.29$, it is about $10^{-4}$. On the other hand, a displacement of the brane away from 
the tip increases this factor.

\section{Relic abundances}  \label{sec:abundance}

In this Section we estimate the relic abundance of the KK modes.
Reheating after brane/anti-brane inflation proceeds through several  stages \cite{lev}.
The brane/anti-brane system mostly annihilates into heavy closed  strings.
The closed strings very quickly decay into the lighter KK modes  \cite{lev,tye2}.
The KK modes are significantly lighter than the closed strings, and  hence they are relativistic when they are produced.
 They then  interact 
among themselves and with the (lighter) SM degrees of  freedom on the brane. We can neglect the small isometry breaking of
  the throat for such
 considerations.
Then, the only KK modes directly coupled to brane fields are those with
zero angular momentum along the directions whose isometries are left
unbroken by the brane. For brevity, we denote such modes by KK$_0$,
and the other KK ones (those that are not directly coupled to the
brane) by KK$_L$. Due to their different behavior, the two types of  modes 
are studied
separately in two Subsections \ref{sub-ab1} and \ref{sub-ab2} below.

From these interactions, one can obtain the relaxation time needed  for the KK modes
to reach thermal equilibrium. It has been estimated in \cite {lev,tye2} that,
for a short (inflationary) throat, the  relaxation time is quicker
than the decay of the KK modes into SM fields. The situation is more  model dependent in the long throat case; in the example
 of a long  throat considered here,
 we fix the parameter ${\cal N}$ by requiring  that the warp factor ``generates'' the electroweak scale at the tip  of the
 throat, cf. Eq.~(\ref{hier}). However, our discussion below is  not constrained 
by  any particular  inflationary mechanism. For instance, it is possible that the 
 SM fields are in the long throat, while the inflationary brane/ anti-brane system is in another throat. In this case, the KK modes 
 and the SM fields
 in the long throat are in thermal equilibrium only  if the tunneling or the decay of the KK modes from the inflationary  to the long 
throat is fast enough;
 this issue is still under  investigation (notice, for example, that different conclusions on the  transfer of this energy density have been reached
 in \cite{tye2} and  \cite{Harling:2007jy}). We will briefly discuss these issues in the last Section.

Our goal is to give a general discussion of the phenomenological  limits due to the long lived KK$_L$ modes. To keep the discussion  as
 general as possible, motivated by the findings of  \cite {lev,tye2}, in the following analysis we simply assume that the KK  modes are
 in thermal equilibrium 
among themselves and with the SM  fields at some given temperature $T$. In this way, their number  density is independent on the history preceding the 
establishment of  the thermal equilibrium, and it is simply
set by the freeze-out temperature at which the interactions changing  their number go out of equilibrium. We show below that, in the region  of parameter 
space of relevance for phenomenology, this freeze out  temperature is typically significantly lower than the mass of the KK  modes, so that their abundance is strongly Boltzmann suppressed.
   As we mentioned, it is an interesting and rather involved problem to estimate the initial number density of KK modes (produced by cascading decays of 
 excited closed strings), and to figure out how far from thermal equilibrium the KK modes will be. If complete thermalization is not achieved, 
one should expect a higher number density for the KK modes, since the Boltzmann suppression will be reduced. For this reason, the phenomenological 
bounds that we  will discuss below based on the assumption of initial thermal  equilibrium should be considered as conservative ones.

\subsection{KK$_0$ modes} \label{sub-ab1}

Let us first study the KK$_0$ modes. We consider both three- and four-legs interactions with the brane fields.
The trilinear interactions have the coupling $\lambda_3$ discussed in Eq.~(\ref{lam3}); it
leads to the decay rate to gauge bosons $V$, (Dirac) fermions $\psi
$, and the Higgs scalar $H$ on the brane \cite{Han:1998sg}
\begin{eqnarray}
\Gamma_{{\rm KK}_0 \rightarrow VV} = \frac{\lambda_{3}^2 \,  m^3}{160 \, \pi}
\;\;\;,\;\;\;
\Gamma_{{\rm KK}_0 \rightarrow {\bar \psi} \, \psi} = \frac{\lambda_ {3}^2 \,
m^3}{320 \, \pi} \;\;\;,\;\;\;
\Gamma_{{\rm KK}_0 \rightarrow H \, H} = \frac{\lambda_{3}^2  \, m^3}{960 \, \pi} \ .
\end{eqnarray}
These expressions assume that the KK modes are much heavier than the
particle they decay into, which is always a good approximations for the
cases we are considering (see below). The complete expressions, with
the kinematical factors included, can be found in \cite{Han:1998sg}.
The total decay rate is
\begin{equation}
\Gamma_{{\rm KK}_0 \rightarrow b \, b}
= \left[ \frac{12}{160 \, \pi} + \frac{3 \times 6 +
3 + 3 \times 1/2}{320 \, \pi} + \frac{1}{960 \, \pi} \right]
\lambda_{3}^2 \, m^3 \simeq 0.047 \, \lambda_{3}^2 \, m^3 \ ,
\label{gtotcomp}
\end{equation}
where $b$ stands for the SM particles at the brane.
For the inverse decay, we instead estimate
\begin{equation}
\Gamma_{b \, b \rightarrow {\rm KK}_0} \sim 10^{-2} \, \lambda_{3}^2 \, N_b \ ,
\end{equation}
where $N_b \sim 0.1 \, T^3$ is the number density of any relativistic  species
on the brane. This estimate is actually
valid for temperatures $T$ comparable or greater than the mass $m$ of
the KK modes. At lower temperatures, the rate has an additional ${\rm
exp } \left( - m / T \right)$ suppression since most of the light
degrees of freedom are not energetic enough to create a KK$_0$ mode
(see for instance the analogous process discussed in appendix B of~
\cite{keith}). 
Assuming that the energy density of the universe is
dominated by the brane degrees of freedom (as we will see, this is
certainly true at temperatures comparable and lower than the masses
of the KK modes), we have the expansion rate
\begin{equation}
H = \frac{\rho^{1/2}}{\sqrt{3} \, M_p} \sim \frac{0.3 \, g_*^{1/2} \,
T^2}{M_p} \ ,
\label{hub}
\end{equation}
where $g_*$ is the number of light degrees of freedom on the brane
(we take $g_* \sim 100 \,$, as it is the case for the standard model
at high temperatures).

This leads to
\begin{eqnarray}
\frac{\Gamma_{{\rm KK}_0 \rightarrow b \, b}}{H} &\sim& 0.17 \left
( \frac{m}{T} \right)^2 \lambda_{3}^2 \, m \, M_p \ , \nonumber\\
\frac{\Gamma_{b \, b \rightarrow {\rm KK}_0}}{H} &\sim& 0.003 \left
( \frac{T}{m} \right) \lambda_{3}^2 \, m \, M_p \;\;\;,\;\;\;
T \gta m \ . 
\label{did}
\end{eqnarray}
where~\footnote{To obtain this quantity we have first used the
expression (\ref{lam3}) for $\lambda_{3}$, (\ref{defxi}) for the mass $m
$ of KK modes, and (\ref{Ms}) for the ratio $V_6 / M_p^2 \,$. The remaining
parameters combine to give one power of ${\cal N} \,$, as defined in
eq.~(\ref{defN}). The coefficient $\xi_{nL}$, has been normalized to
the typical value assumed by the lightest KK modes in the spectrum.}
\begin{equation}
\lambda_{3}^2 \, m \, M_p \simeq 220 \, {\cal N} \, \left
( \frac{\xi_{nL}}{5} \right) \left( \frac{R}{\sqrt{\alpha'}} \right)^{-7} \ .
\end{equation}

We see that both the decay and the inverse decay are effective (namely, $\Gamma > H$) 
when $T \simeq m \,$ if $R / \sqrt{\alpha'} \lta 11$ for the short throat  case, and $R / \sqrt{\alpha'}
\lta 250$ for the long throat case.

The four-legs interactions are proportional to the coupling constant 
$\lambda_{4} = \lambda_3^2$ given in Eq.~(\ref{lam4}). It is instructive to compare
the rates of the four- and three-legs interactions. For $T \gta m$, we
find~\footnote{For the $2 \rightarrow 2$ interactions, we assume
that the $s$-wave process is unsuppressed; for the present
scattering, we have actually computed the cross sections for the
transverse--traceless helicity of the KK$_0$ modes. The cross section
for the other helicities is increased by additional powers of $(T / m)
$ in the $T > m$ regime.}
\begin{eqnarray}
\frac{\Gamma_{{\rm KK}_0 \, {\rm KK}_0 \rightarrow {\rm b} \, {\rm
b}}}{\Gamma_{{\rm KK}_0 \rightarrow {\rm b} \, {\rm b}}} &\sim& \frac
{10^{-2} \, \lambda_{4}^2 \, T^2 \, N_{{\rm KK}_0} \, g_*}{10^{-2} \,
\lambda_{3}^2 \, m^3 \, g_*} \sim 0.5 \, \lambda_{3}^2 \, m^2 \, \left( \frac{T}{m} \right)^5
\;\;,\;\; T \gta m \ , \nonumber\\
\frac{\Gamma_{b \, b \rightarrow {\rm KK}_0 \, {\rm KK}_0}}{\Gamma_
{ b \, b \rightarrow {\rm KK}_0}}
&\sim& \frac{10^{-2} \; \lambda_{4}^2 \, T^2 \, N_b}{10^{-2}
\; \lambda_{3}^2 \, N_b} \sim
\lambda_{3}^2 m^2 \left( \frac{T}{m} \right)^2 \;\;,\;\; T \gta m \ ,
\label{4vs3}
\end{eqnarray}
where
\begin{equation}
\lambda_{3}^2 \, m^2 \simeq 5 \times 10^4 \left( \frac{\xi_{nL}}
{5} \right)^2 \left(
\frac{R}{\sqrt{\alpha'}} \right)^{-8} \ .
\end{equation}

In general, the $2 \rightarrow 2$ scatterings dominate at high
temperatures, and the moment at which the trilinear interactions
become dominant depends on the value of $R / \sqrt{\alpha'} \,$. One  can also
check that, at the temperature $T \simeq m$, these rates are also  greater than the
expansion rate (\ref{hub}), for $R / \sqrt{\alpha'} \lta 6$ in the  short throat case,
and for $R / \sqrt{\alpha'} \lta 37$ in the long throat case. If this  is the case,
such interactions lead to thermal  equilibrium between the KK$_0$ modes
and the brane fields. 

As the temperature drops below the mass of a KK$_0$ mode, both the
ratios (\ref{4vs3}) acquire an ${\rm exp } \left( - m / T \right)$
suppression factor (in this first case this is due to the Boltzmann
suppression of $N_{{\rm KK}_0}$; in the second case this is due to
the fact that the interaction at the numerator produces one more
heavy particle than the one at the denominator). Also the inverse
decay is exponentially suppressed in this regime, as we have  discussed above.
Therefore, the only
interaction that remains active in the nonrelativistic regime is the
decay of the KK$_0$, which quickly eliminates these modes.

\subsection{KK$_L$ modes} 
\label{sub-ab2}

The decay rate of the angular KK$_L$ modes into brane fields is suppressed by the 
 small isometry breaking of the throat. The decay rate is still given
  by Eq.~(\ref{gtotcomp}), but now the coupling $\lambda_{3}$ is  given by Eq.~(\ref{lameps}). Assuming 
that the brane is at the tip of  the throat ($y_b = y_t$),
and using the relations (\ref{Ms}), (\ref{defN}), and (\ref{defxi}),  we can
rewrite the total rate normalized by the KK mass as
\begin{equation}
\frac{\Gamma_{{\rm KK}_L \rightarrow b \, b}}{m} \simeq \frac{90 \,  \xi_{nL}^2}{N^{2
\alpha}} \,
\left( \frac{\sqrt{\alpha'}}{R} \right)^{8 - 6 \alpha} \, \left( \frac
{V_6}{R^6} \right)^\alpha \ .
\label{gtot}
\end{equation}

Let us now estimate the relic abundance of these modes.
Under the assumption of isometric throat, the KK$_L$ modes are not  coupled
to the brane fields. However, they interact with the KK$_0$ modes studied
in the previous subsection. As long as the modes are relativistic, we
have~\footnote{Also in this case, the estimate for the $2 \rightarrow
2$ scattering is given for the transverse-transpose helicity;
scatterings between different helicities have higher powers of $T/m$.}
\begin{eqnarray}
T \gta m \; : \;\; \left\{ \begin{array}{l}
\Gamma_{{\rm KK} \rightarrow {\rm KK} \, {\rm KK}} \;\;\sim\;  10^{-2}
\, \lambda_{3}^2 \, m^3  \\ \\
\Gamma_{{\rm KK} \, {\rm KK} \rightarrow {\rm KK}} \;\;\sim\; 10^{-2}
\, \lambda_{3}^2 N_{\rm KK}
\\ \\
\Gamma_{{\rm KK} \, {\rm KK} \rightarrow {\rm KK} \, {\rm KK}} \sim\;
10^{-2} \, \lambda_{4}^2 N_{\rm KK} \, T^2  \ . \end{array} \right.
\end{eqnarray}

We do not need to distinguish between KK$_0$ and KK$_L$ modes in
these estimates. Analogously to what we saw in the previous
subsection, the $2 \rightarrow 2$ scatterings are typically the
dominant interactions in the relativistic regime. Also in this case,
the rate for such interactions is faster than the Hubble rate. Indeed
\begin{equation}
\frac{\Gamma_{{\rm KK} \, {\rm KK} \rightarrow {\rm KK} \, {\rm KK}} }
{H} \sim 2 \times 10^4 \left( \frac{\xi_{nL}}{5} \right)^3 \, {\cal N}
\, \left( \frac{R}{\sqrt{\alpha'}} \right)^{-15} \, \left( \frac{T}
{m} \right)^3
\;\;\;,\;\;\; T \gta m \ ,
\end{equation}
which is greater than one for $T > m \,$ and for $R / \sqrt{\alpha'}  \lta 4.6$
in the short throat case, and for $R / \sqrt{\alpha'} \lta 26$ in the  long throat case.
In the following, we assume that this is the case, so that also the KK $_L
$ modes are in thermal equilibrium with the brane fields in this regime.

These reactions slow down and eventually freeze out when the KK$_L$
modes become nonrelativistic. In this regime there are three relevant
interactions that can potentially reduce the abundance of a KK$_L$
mode. The first is a $2 \rightarrow 2$ scattering with brane degrees
of freedom, mediated by a virtual KK$_0$ mode. The second is a $2
\rightarrow 2$ scattering into two KK$_0$ modes.  The third is a $2
\rightarrow 1$ inverse decay, producing a KK$_0$ mode. Once formed,
the KK$_0$ modes then quickly decay into the brane degrees of freedom (as
we saw in the previous subsection). In all these cases, the two
incoming KK$_L$ modes must combine to produce a zero angular
momentum (in the directions corresponding to the isometries unbroken
by the brane). This is also the reason why a single KK$_L$ mode
cannot interact with KK$_0$ or the brane modes.

The rate for these interactions can be roughly estimated as
\begin{eqnarray}
T \lta m \; : \;\; \left\{ \begin{array}{l}
\Gamma_{{\rm KK}_L \, {\rm KK}_{-L} \rightarrow b \, b} \;\;\;\;\;\;
\sim\;  10^{-2} \, \lambda_{3}^4 \,
N_{{\rm KK}_L} \, m^2 \, g_*  \\ \\
\Gamma_{{\rm KK}_L \, {\rm KK}_{-L} \rightarrow {\rm KK}_0 \, {\rm
KK}_0} \sim\; 10^{-2} \; \lambda_{4}^2 \; N_{{\rm KK}_L} \,
m^2 \\ \\
\Gamma_{{\rm KK}_L \, {\rm KK}_{-L} \rightarrow {\rm KK}_0} \;\;\;\;
\sim\; 10^{-2} \, \lambda_{3}^2 \, N_{{\rm KK}_L} \;\;\;,\;\;
\; {\rm if \; sufficiently \; high \; momentum}
\end{array} \right.
\label{ratestim}
\end{eqnarray}

Since $\lambda_{4} \approx \lambda_{3}^2 \,$, the first
interaction dominates over the second one, due to the greater amount
of available channels (this is certainly true for
the lightest KK$_L$ modes, so that there are few lighter
KK$_0$ modes they can annihilate into; as we will see, the abundance of
the heavier KK$_L$ modes is anyhow suppressed relative to the
lighter ones). The last process can in principle be faster but (due
to kinematical reasons) it typically involves only a negligible
amount of KK$_L$ particles. The two incoming KK$_L$ particles can
inverse decay only into a heavier KK$_0$ species. More precisely,
denoting by $m_0$ the mass of the KK$_0$ mode, and by $m$ the one of
the incoming KK$_L$ particles of a given species, the momentum of
the two particles in the center of mass frame is forced to be
\begin{equation}
k = \sqrt{ \frac{m_0^2}{4} - m^2 } \ .
\label{momentum}
\end{equation}
As we saw, the masses of the KK modes are discretized in units of
\begin{equation}
\Delta m \sim \frac{\pi}{R} \, {\rm e}^{-y_t / R} \ ..
\label{deltam}
\end{equation}
It may be possible that, for some special case, there is a KK$_0$
mode having a mass $m_0$ slightly greater but very close to $2 \, m \,
$. In general, however, the value in (\ref{momentum}) will be of the
order of $\Delta m$. For the lightest among the KK$_L$ modes, $
\Delta m \sim m \,$. In the nonrelativistic regime, the abundance of
the KK$_L$ mode that we are considering is Boltzmann suppressed.  Most of
such particles have momentum $k \sim T \ll m \,$. The inverse decay $
{\rm KK}_L \, {\rm KK}_{-L} \rightarrow {\rm KK}_0$ can only happen
if, in the center of mass frame, the particles have the much higher
momentum (\ref{momentum}).
So, only a very few of the KK$_L$ modes will be able to perform the
inverse decay. The remaining particles do not have enough energy, and
can be depleted only
through the first two interactions in~(\ref{ratestim}).

For this reason, we neglect the effect of the inverse decay in the
abundance of the lightest KK$_L$ modes 
(in this, we differ from \cite{tye2}, where the suppression
that we have just discussed
was not considered). We therefore need to
estimate the moment at which the first process in~(\ref{ratestim})
freezes out. From this, we then obtain the relic abundance
of the KK$_L$ modes. In the nonrelativistic regime,
\begin{equation}
N_{{\rm KK}_L} \simeq 5 \, \left( \frac{m \, T}{ 2 \pi } \right)^
{3/2} \, {\rm e}^{-m/T} \ .
\label{nkpm}
\end{equation}
Using the expression~(\ref{hub}) for the Hubble rate we find (see the
previous subsection for more details in obtaining the final estimate)
\begin{eqnarray}
\frac{\Gamma_{{\rm KK}_L \, {\rm KK}_{-L} \rightarrow {\rm b} \,
{\rm b}}}{H} &\sim&
0.1 \, \lambda_{3}^4 \, m^3 \, M_p \, \left( \frac{m}{T} \right)
^{1/2} \, {\rm e}^{-m/T} \nonumber \\
&\sim& 10^6 \, {\cal N} \, \left( \frac{\xi_{nL}}{5} \right)^3 \, 
\left( \frac{\sqrt{\alpha'}}{R} \right)^{15} \, \left( \frac{m}{T} \right)^
{1/2} \, {\rm e}^{-m/T} \ .
\label{freezeout}
\end{eqnarray}
Equalizing this to unity, on can estimate the freeze-out temperature $T_f$.

\begin{figure}[h]
\centerline{
\includegraphics[width=0.4\textwidth,angle=-90]{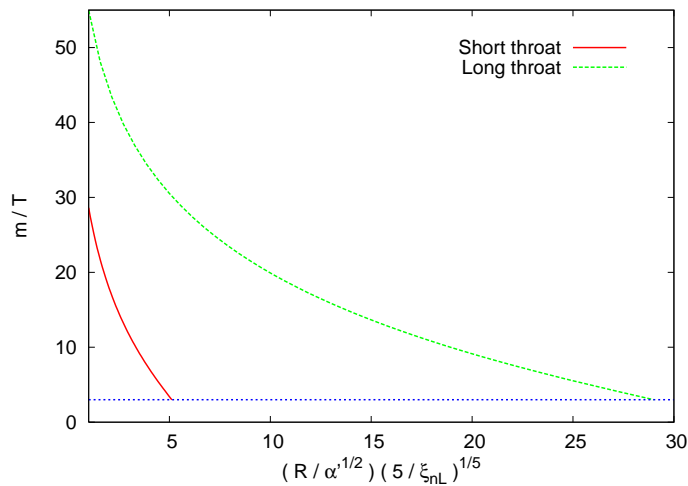}
}
\caption{Estimate for the $m / T$ ratio at which the scatterings $
{\rm KK}_L + {\rm KK}_{-L} \rightarrow b + b$ freeze out. The value
is strongly sensitive to the ratio $R/\sqrt{\alpha'}$.}
\label{fig-mt}
\end{figure}

In Figure~\ref{fig-mt} we show the ratio of $m / T$ when the
scatterings go out of equilibrium in the case of parameters which are
typical of either a short or a long throat. We also set a rough bound
of $m / T \gta 3$ for the nonrelativistic approximation for the KK
modes to be valid~\cite{kolbturner}. For higher values of $R/\sqrt
{\alpha'}$ the KK$_L$ decouple while still relativistic, which results in a greater
abundance (at even higher values of $R / \sqrt{\alpha'}$ we could have
that the scattering of the KK modes are never in thermal  equilibrium; the
abundance of the KK particles then depends on the specific details  of the reheating scenario;
due to this strong model dependence, we do not consider such cases  here).

\begin{figure}[h]
\centerline{
\includegraphics[width=0.4\textwidth,angle=-90]{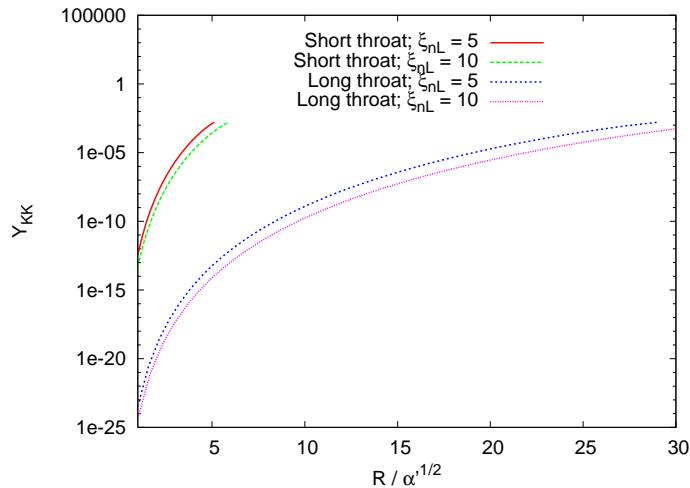}
}
\caption{Relic abundance for the KK$_L$ species as function of parameter $R/\sqrt{\alpha'}$.
We note that the lightest modes are more abundant than the heavier ones.
}
\label{fig-abu}
\end{figure}

Since this is the fastest interaction decreasing the number of KK$_L$, 
the moment at which this interaction freezes out sets the relic
abundance of this species. In general, the later the freeze out
occurs, the lower is the relic abundance of the mode (since the
Boltzmann suppression is greater). From the horizontal axis of the Figure~\ref{fig-mt}
 we see that, for any fixed value of $R / \sqrt{\alpha'}
$, the higher the mass of a KK$_0$ mode (greater $\xi_{nL}$), the
higher is the value of $m/T$ at which the freeze out occurs (later
time). Therefore, the KK$_L$ modes with higher mass end up with a
lower relic abundance. We show this in Figure~\ref{fig-abu}, where we  plot the
ratio $Y_{{\rm KK}_L} \equiv N_{{\rm KK}_L} / s$ between the
resulting number density $N_{{\rm KK}_L}$ and the entropy of the
relativistic fields on the brane
\begin{equation}
s \simeq \frac{2 \pi^2}{45} g_* \, T^3
\end{equation}
for the two cases $\xi_{nL} = 5 ,\, 10 \,$. Also in this plot, we
show only the cases for which the KK$_L$ decouple while
nonrelativistic.

We can verify that the energy density of
the KK$_L$ modes is negligible with respect to that of the brane
fields at the time at which they decouple (we used this assumption
in setting the Hubble rate~(\ref{hub})). Since the KK$_L$ are
nonrelativistic, their energy density is simply their number density
times their mass. For  SM radiation (i.e. the light degrees of freedom on the brane), the
energy density is instead related to the entropy density through $
\rho = (3 \, T / 4) s \,$. This gives
\begin{equation}
\frac{\rho_{\rm KK_L}}{\rho_{\rm rad}} \simeq \frac{4}{3} \left
( \frac{m}{T} \right) \, Y_{{\rm KK}_L} \ll 1 \ .
\label{rhorho}
\end{equation}

It is also worth noting that, for the specific case of warped brane/anti-brane inflation, the temperature after inflation is never 
higher than the mass of the KK modes. Indeed, even assuming that the thermalization occurs on a very quick timescale, 
the energy density during inflation, $\rho \simeq {\rm e}^{-4 y_t / R} / (4\pi^3 g_s \alpha'^2)$, leads to  a maximal 
reheating temperature $T_{\rm max} \sim 0.23 \, {\rm e}^{-y_t / R} / \sqrt{\alpha'} \,$.  Using the expression (\ref{defxi}) 
for the KK mass, we have
\begin{equation}
\frac{m}{T_{\rm max}} \sim 22 \left( \frac{\xi_{nL}}{5} \right) \left ( \frac{\sqrt{\alpha'}}{R} \right) \ .
\end{equation}
As we can see in the Fig.~\ref{fig-abu}, this value is comparable with  the freeze-out temperature that we have obtained above. 
Therefore, our estimate for the number density is reliable in this case only if the thermalization is quick, 
as it appears from the estimates of the relaxation time given in \cite{lev,tye2}.

\section{Phenomenological Constraints} 
\label{sec:constraints}

We are interested in the cosmological effects of the KK$_L$ modes.
Such modes are long-lived, since
their decay rate is proportional to the small isometry breaking of
the throat. We start by expressing the mass of the KK modes  $m_n$ in a more
explicit form. By combining the three relations (\ref{Ms}), (\ref
{defN}), and (\ref{defxi}), we can write
\begin{equation}
\frac{m_n}{M_p} \simeq \frac{44}{\cal N} \, \xi_{nL} \left( \frac{\sqrt{\alpha'}}{R} \right) \ .
\label{mnmp}
\end{equation}
As we have seen in the previous Section, the lightest KK modes are in
general those with the greatest abundance. Therefore, for
definiteness, in the remainder of this analysis we consider only the
presence of a single species with $\xi_{nL} = 5 \,$, as it is typical
for the lightest modes (stronger effects, and more stringent bounds,
are obtained if more $KK_L$ are generated with a comparable
abundance). The typical values for ${\cal N}$ are given in Eqs.~(\ref
{Ithroat}) and (\ref{SMthroat}). In the range of $R / \sqrt{\alpha'}$
considered in figures \ref{fig-mt} and \ref{fig-abu} above, we find
\begin{eqnarray}
m &\sim& \left( 2 \times 10^{14} - 10^{15} \right) {\rm GeV} \;\;\;\;\;
\;,\;\;\;\;\;\;
{\rm short \; throat} \nonumber\\
m &\sim& \left( 200 - 5,000 \right) {\rm GeV} \;\;\;\;\;\;
\;\;\;\;,\;\;\;\;\;\;\;
{\rm long \; throat}
\label{massrange}
\end{eqnarray}

\begin{figure}[h]
\centerline{
\includegraphics[width=0.4\textwidth,angle=-90]{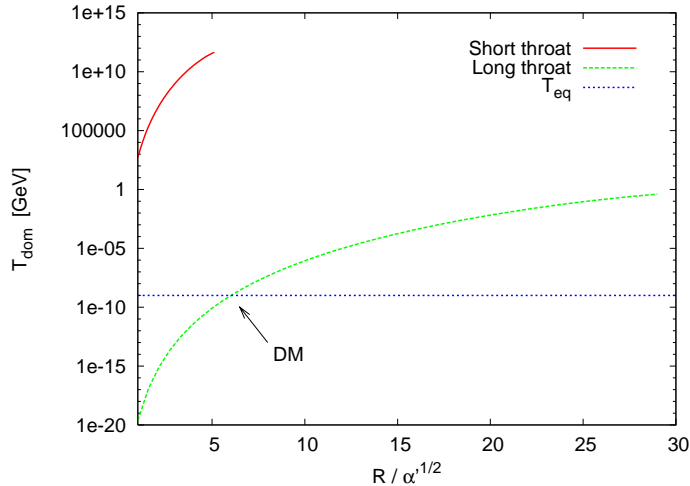}
}
\caption{
Temperature of the  universe at which the relic KK$_L$ particles start to dominate,
(provided they have not decayed yet) as function of $R/\sqrt{\alpha'}$.
 The temperature $T_{eq} \simeq 7.4 \cdot 10^{-10} \, {\rm GeV} \,$ at the moment of matter--radiation
equality is also shown for comparison. In the long throat, for $R / \sqrt{\alpha'} \simeq 6 \,$,
the KK modes can be the dark matter candidate.
}
\label{fig-tdom}
\end{figure}

As Eq.~(\ref{rhorho}) shows, the KK$_L$ particles dominate the
energy density of the universe at sufficiently low temperature,
provided they have not decayed before that. From the value of $Y_
{{\rm KK}_L} \,$ shown in Fig~\ref{fig-abu}, and from the expression~
(\ref{defxi}) for the mass of the KK particles, we can find the
temperature of the SM radiation,  denoted by $T_{\rm dom} \,$,
 at which the KK particles
start to dominate\footnote{There is actually a slight difference
in the value of $g_*$ entering in the formula for the energy density
and the entropy after $e^{\pm}$ annihilation \cite{kolbturner}. This
introduces a factor $1.16$ in Eq.~(\ref{rhorho}) for $T \lta 0.5 \,
{\rm MeV} \,$. This effect is negligible at the level of accuracy of
the present estimates.}. The value  $T_{\rm dom} \,$ as function of the parameters 
is shown in Figure~\ref{fig-tdom}; for comparison, we also show the  value
of the CMB temperature $T_{\rm eq} \simeq 7.4 \cdot 10^{-10} \, {\rm
GeV}$ when matter starts to dominate  in the  universe~\footnote
{Cosmological quantities in our universe
are obtained by using the WMAP values $\Omega_\Lambda \simeq 0.73 ,\,
\Omega_m \simeq 0.27 ,\, h \simeq 0.70 \,$ \cite{WMAP}.}.

Notice that intersection of two curves, $T_{eq}$ and  $T_{\rm dom} \,$ for the long throat selects the value of the parameter
$R/\sqrt{\alpha'} \simeq 6$, in which case the KK modes can be the dark matter candidate. 
As we will see below, this also poses restrictions on the possible values of $\alpha \,$. The Figure~\ref{fig-tdom} will be  discussed in details in the next two Subsections,
where the two types of throats are studied separately.

\subsection{Short throat}

It is clear from Fig.~\ref{fig-tdom} that, for the case of a short  throat, it is phenomenologically required that
all the KK$_L$ modes decay. If this not the case, our universe  would have become (and remained) matter dominated much before $z_{eq} \sim 10^4$,
the redshift of  matter-radiation equality,  indicated by the
 observed values of $\Omega_m$ and of the CMB  temperature (in other words, $\Omega_m$ today would be too big).

 We therefore assume that the KK$_L$ particles decay  with the rate~(\ref{gtot}). We treat $\alpha$ as a free parameter.
  Typically, smaller values of $\alpha$ correspond to a greater break  of isometry of the throat, and, consequently, to a quicker decay.
  From Eqs.~(\ref{gtot}) and (\ref{mnmp}),
 we find the lifetime of the angular KK modes
\begin{equation}
\tau \simeq \frac{1}{\Gamma_{\rm tot}} \simeq 1.1 \times 10^{-27} \,
 {\rm s} \, \left( 4 \times 10^3 \right)^{2
 \left( \alpha - 1.29 \right)} \, \left( \frac {R}{5 \, \sqrt{\alpha'}} \right)^{9 - 6 \alpha} \,
 \left( \frac{R}{V_6^ {1/6}} \right)^{6 \alpha} \ .
\label{tdecashort}
\end{equation}

\begin{figure}[h]
\centerline{
\includegraphics[width=0.4\textwidth,angle=-90]{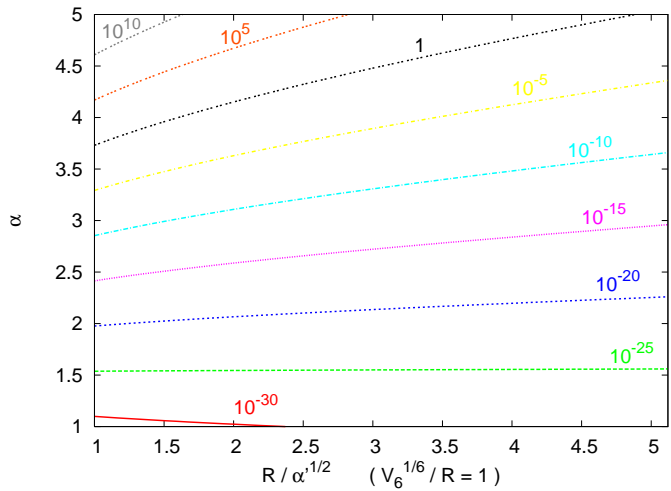}
}
\caption{Curves of 
lifetime (labelled by  seconds) of the KK$_L$ particles in the short throat  case, as function of $\alpha$ and $R/\sqrt{\alpha'}$.
 For definiteness, we have fixed $V_6^{1/6} / R = 1$ in this  plot. Higher values 
of this ratio correspond to a shorter lifetime,  cf. Eq.~(\ref{tdecashort}).
}
\label{fig-lifeshort}
\end{figure}

We see that this lifetime is strongly dependent on the parameters of  the model.
 For illustration, we show it in Figure~\ref{fig-lifeshort}  for a fixed value of $V_6^{1/6} / R$.

Unless $\alpha$ is very high (very long lifetimes),
the strongest phenomenological limits are those related to the effect  that
the decay products have on the abundance of light elements formed at
Big--Bang Nucleosynthesis (BBN). Different limits are obtained for
radiative and hadronic decays. KK$_L$ have both type of decays,
with  ${\rm O } \left( 1 \right)$ branching ratios \cite{Han:1998sg}.
If we separate in Eq.~(\ref{gtotcomp}) the rates into quarks and
gluons from the other particles, we find a hadronic branching ratio
of $0.73 \,$. Although the radiative decays are those which have been 
traditionally more studied, hadronic decays lead to stronger bounds.
The BBN limits for a pure hadronic decay can be found in
the Figures 38, 39, and 40 of \cite{moroi}. These three figures present
the constraints for decaying
particles of mass $m = 100 ,\, 1,000 ,\, 10,000 \,{\rm GeV} \,$,
respectively, in the $\tau - m \, Y$ plane (for any given lifetime $
\tau$, there is an upper limit on the allowed values $m \, Y$). The
three cases do not show substantial differences from each other (at
least, not at the level of accuracy of the present computation).

The limits of \cite{moroi} have been approximately reproduced
in our Fig.~\ref{fig-limit} below. We notice that the BBN limit is  strongly relaxed and eventually
disappears for lifetimes shorter than $\sim 10^{-2} \, {\rm s} \,$:  if a particles decays
before this time, the decay products thermalize before BBN starts.
In our case, the product $m Y$ is an
increasing function of $R /\sqrt{\alpha'}$, while it is independent of
the other ``free parameters'' of the model ($V_6^{1/6} / R$ and $ \alpha \,$). For
the values of $R /\sqrt{\alpha'}$ we are considering, $m \, Y$ ranges  from
$\sim 400 \, {\rm GeV}$ to $\sim 3 \times 10^{11} \, {\rm GeV} \,$.
Such values are sufficiently high so that consistency with BBN  requires that the decay
took place place before $\sim 10^{-2} \, {\rm s} \,$~\footnote{It  should
be noted that the results of~\cite{moroi} have been obtained for much  smaller masses than the typical KK masses for
 the short throat;  nonetheless, we see that for all the masses considered in \cite {moroi} (cf. their Figures 38, 39, and 40),
 the BBN bound disappear  at $\tau \lta 10^{-2} \, {\rm s} \,$. For this reason, we take such  value as a rough bound in the case
 of the higher injection energy  considered here.}.

\begin{figure}[h]
\centerline{
\includegraphics[width=0.4\textwidth,angle=-90]{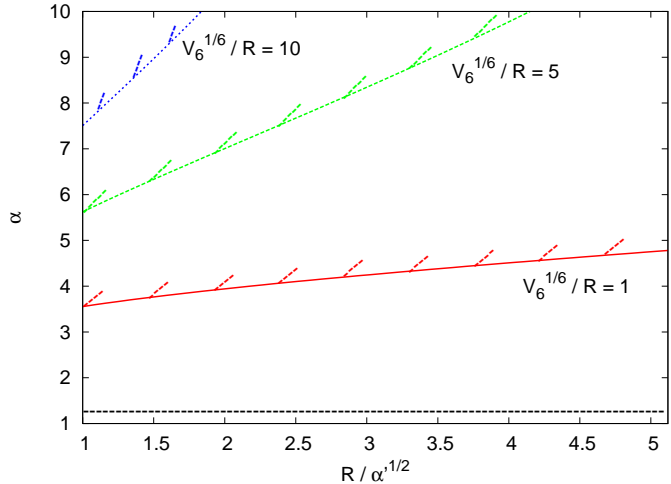}
}
\caption{
BBN limit $\tau < 10^{-2} \, {\rm s}$ for three reference values of  the ratio $V_6^{1/6} / R$ for the  short throat. 
 For each case, regions  above the lines (greater lifetime)  are excluded.
  The horizontal line at $\alpha = 1.29$ corresponds to the reference value.}
\label{fig-bbnshort}
\end{figure}

In Figure~\ref{fig-bbnshort} we show the values of parameters leading  to $\tau = 10^{-2} \, {\rm s} \,$.
 The three lines refer two three  reference values of the ratio $V_6^{1/6} / R$. For each case, values  above
 the lines shown result in a greater lifetime, and are therefore  excluded.

It is possible that the KK$_L$ particles dominate before decaying.  This gives rise to an intermediate stage of
 matter domination. As  long as this stage takes place before BBN, such a possibility is  phenomenologically 
allowed~\footnote{Such stage can have interesting consequences
 for cosmology; for instance, it affects the  relation between the  wavelengths  of observable cosmological perturbations 
and  the moment when this mode 
leaves the horizon (efolds number)  during inflation.}.  To see whether this is  the
 case, we compute the temperature of the thermal bath on the brane  when the decay takes place; by equating the decay rate (\ref{gtot})
  (with the expression~(\ref{mnmp}) for the mass) to the Hubble rate~(\ref{hub}), we find
\begin{equation}
T_{\rm decay} \simeq 2.2 \times 10^{10} \, {\rm GeV} \, \left( 4 \times  10^3  \right)^{1.29-\alpha}
\, \left( \frac{100} {g_*} \right)^{1/4} \left( \frac{5 \, \sqrt{\alpha'}} {R} \right)^{9/2 - 3 \alpha}
\left( \frac{V_6^{1/6}}{R} \right)^{3 \alpha} \ .
\label{temdeca}
\end{equation} 
In Fig.~\ref{fig-tdom} above we showed the temperature $T_{\rm dom}$  of the brane at  which the KK$_L$ particles dominate, if they have
  not decayed yet. Therefore if the value of $T_{\rm decay}$ that we  have just found is greater than the one of $T_{\rm dom}$ shown in 
 that figure, the KK$_L$ modes decay before dominating; the decay  increases the temperature of the thermal bath by a negligible amount.
  If, on the contrary, $T_{\rm dom} > T_{\rm decay}$, the KK$_L$  dominate the energy density of the universe when they decay. In this
  case, $T_{\rm decay}$ actually refers to the temperature of the  thermal bath formed by the decay products~\footnote{The  thermalization
 of the decay products takes place on a much quicker  timescale than the decay of the KK$_L$ particles, and can therefore  be considered
 as instantaneous for this discussion.}, which dominates  over the pre-existing one on the brane.

\begin{figure}[h]
\centerline{
\includegraphics[width=0.4\textwidth,angle=-90]{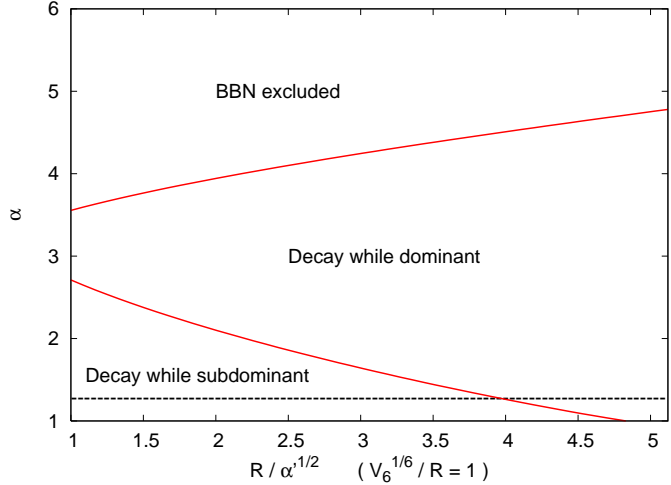}
}
\caption{
Different scenarios  for the decay of the KK$_L$ particles in  the short throat, for the fixed value
$V_6^{1/6} / R = 1 \,$.   The horizontal line corresponds to the reference value $\alpha = 1.29$.}
\label{fig-regshort}
\end{figure}

We show this effect in Figure \ref{fig-regshort}, for the specific  case of $V_6^{1/6} / R = 1 \,$. The upper line is also shown 
in Fig.~\ref{fig-bbnshort}. Points above this line lead to a lifetime $> 10^ {-2} \, {\rm s} \,$, and conflict with BBN bounds. 
For points in the  intermediate region, the KK$_L$ particles dominate the energy  density of the universe when the decay. 
Prior to the decay, the  universe is therefore matter dominated.

\subsection{Long throat}

Proceeding as in the previous Subsection, we now find the lifetime  for the KK$_L$ particles
\begin{equation}
\tau \simeq \frac{1}{\Gamma_{\rm tot}} \simeq 3 \times 10^{13} \,
 {\rm s} \, \left( 8 \times 10^{14} \right)^{2
 \left( \alpha - 1.29 \right)} \, \left( \frac {R}{5 \, \sqrt{\alpha'}} \right)^{9 - 6 \alpha} \,
 \left( \frac{R}{V_6^ {1/6}} \right)^{6 \alpha}
\label{tdecalong} \ ..
\end{equation}

\begin{figure}[h]
\centerline{
\includegraphics[width=0.4\textwidth,angle=-90]{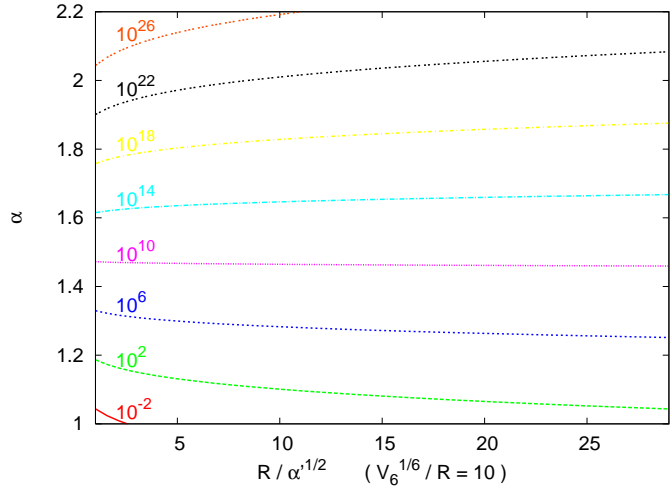}
}
\caption{
Lifetime (in seconds) of the KK$_L$ particles in the long throat.  For definiteness, we have fixed
$V_6^{1/6} / R = 10$ in this plot. Higher (shorter) values of this  ratio correspond to a shorter (longer) lifetime, cf. Eq.~(\ref {tdecalong}).}
\label{fig-longlife}
\end{figure}

The lifetime is even more sensitive to the parameters of the model  (particularly, to $\alpha$) than in the previous case. For
  illustration, we show it in Figure~\ref{fig-longlife} for the  specific value $V_6^{1/6} / R = 10 \,$ (as clear from Eq.~(\ref {tdecalong}), 
the lifetime is a decreasing function of this  quantity). We see that the lifetime changes by several order of  magnitudes in the range
 of value for $\alpha$ shown (notice that this  range is smaller than the one shown in Fig.~\ref{fig-lifeshort} for  the short throat case).

We also observe that the lifetime of the modes can be easily  comparable or much greater than the present age of the universe 
($t_0  \simeq 4.3 \times 10^{17} \, {\rm s}$). The value of $\alpha$ at which
the two timescales are equal is only weakly (logarithmically)  dependent on the other two ``free parameters'' of the model; for 
 instance we find $\alpha \sim 1.4$ for $V_6^{1/6} / R = 1$, and for  all the values of $R / \sqrt{\alpha'}$ considered in the 
long throat  case; this is not much different than the value $\alpha \sim 1.8$ that can be seen in Fig.~\ref{fig-longlife} 
for the $V_6^{1/6} / R = 10$  case. Due to the strong dependence of the lifetime on $\alpha$,  slightly higher values of this
 parameter result in a lifetime which  is much longer than the present age of the universe.

\begin{figure}[h]
\centerline{
\includegraphics[width=0.4\textwidth,angle=-90]{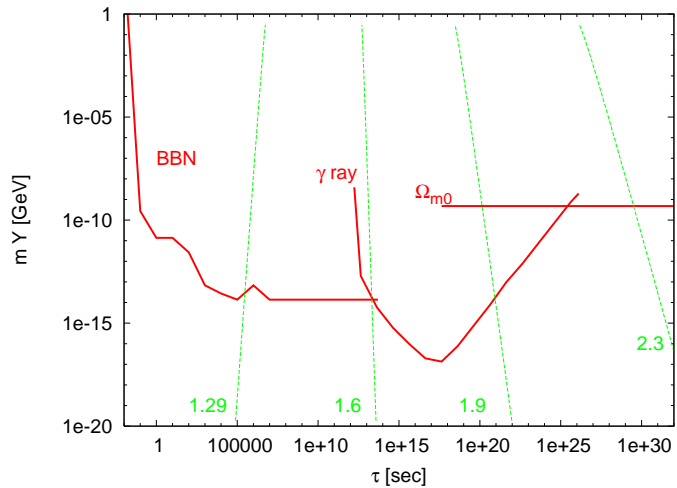}
}
\caption{
Combined upper bound on the mass $m$ times the abundance $Y$ for an  unstable particle with lifetime $\tau$.
 The BBN limit is taken from  \cite{moroi}, while the bound from the diffuse $\gamma$ ray  background from
 \cite{diffusegamma}. The horizontal line denoted by $ \Omega_{m0}$ is the limit imposed by requiring that the 
energy density  of the KK particle does not exceed  the   dark matter $\Omega$.
The other curves represent the values of $\tau$ and $m \, Y$ obtained  in the long throat, for $V_6^{1/6}
 / R = 10 \,$, for different values of $\alpha$ (indicated on the  lines) and of $R / \sqrt{\alpha'}$
($m Y$ is an increasing function of $R / \sqrt{\alpha'}$).
}
\label{fig-limit}
\end{figure}

In Figure~\ref{fig-limit} we show the phenomenological limits on an  unstable particle of mass in the range
(\ref{massrange}) for the long throat case. Limits from BBN are taken  from \cite{moroi} (see the discussion in the previous subsection), 
 while the ones from the diffuse $\gamma$ ray background are taken  from Fig.~12 of \cite{diffusegamma}. In both cases, the strongest 
 limit comes from hadronic decays, which, for the KK$_L$ particles,  have a branching ratio of $\sim 0.73 \,$. In both cases, the
  phenomenological limits show a weak dependence on the mass of the  particle in the interval of our interest; this dependence is 
 negligible at the level of accuracy of the present analysis. We also  show the limit obtained by requiring that the energy 
density of the KK $_L$ particles is smaller than that of dark matter in our universe  (this limit is relevant only for lifetimes greater
 than the present  age of the universe).

Such limits have to be compared with the values of $\tau$ and of $m  \, Y$ obtained for our model. For illustrative purposes, we 
also show  in Fig.~\ref{fig-limit} the values obtained in the specific case of  $V_6^{1/6} / R = 10$ (same choice as in 
 Fig.~\ref{fig-longlife}).  Each line shown corresponds to a given value of $\alpha$, while $R /  \sqrt{\alpha'}$ 
ranges in the interval $\left[ 1 , 29 \right]$ (that  is, the usual one for the long throat case). As we already mentioned,
  the lifetime strongly increases with $\alpha$.

\begin{figure}[h]
\centerline{
\includegraphics[width=0.4\textwidth,angle=-90]{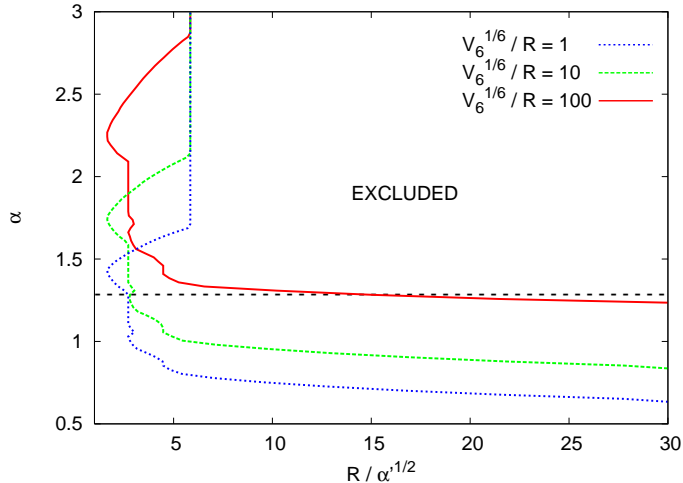}
}
\caption{
Final exclusion region for the parameter in the long throat. The  three lines correspond to three reference value of $V_6^{1/6} / R$. 
 For each case, values of the parameters on the right of the  corresponding curve conflict with the phenomenological limits shown 
 in Fig.~\ref{fig-limit}. The highest values of $\alpha$ shown result  in KK$_L$ particles with a much longer lifetime than the 
age of the  universe. In this case,  the only relevant bound is that the energy  density of the KK$_L$ particles does not exceed the
 one of dark  matter in our universe.    The horizontal line corresponds to  $\alpha = 1.29$}.
`\label{fig-long}
\end{figure}

The quantity $m \, Y$ only depends on $R / \sqrt{\alpha'} \,$, and it  is a growing function of this ratio. Therefore, for any fixed values 
 of the other ``free parameters'' of the model ($V_6^{1/6} / R$ and $ \alpha$), only values of $R / \sqrt{\alpha'} \,$ smaller than some  given
 quantity are phenomenologically acceptable. We show this in  Fig.~\ref{fig-long}. The three lines shown correspond to three  reference values 
of $V_6^{1/6} / R$. For all the cases, the points on  the right of the corresponding line are phenomenologically excluded.  For
 values of $R / \sqrt{\alpha'}$ saturating this bound, the KK$_L$ particles constitute the dark matter of our universe. In each case, 
 the limit on $R / \sqrt{\alpha'} \,$ is strongest for an intermediate  value of $\alpha$; this is the value for which the lifetime of
 the KK $_L$ particles is comparable to the present age of the universe (we  see in Fig.~\ref{fig-limit} that this is the point at which $m \, Y$  is 
most constrained). At higher values of $\alpha$, the lifetime of  the KK$_L$ particles becomes much greater than the present age of  the universe.
 In this case the only relevant bound is that their  energy density does not exceed the one of dark matter in the  universe. Since the energy density
 depends only on $R / \sqrt{\alpha'} $, we find the same limit $R / \sqrt{\alpha'} \lta 6$ for all the  values of $\alpha$ and $V_6^{1/6} / R$ in this range.  
  When this limit is saturated, the KK modes are identified with the dark matter of the universe. 
On the contrary, 
 for the lowest values of $\alpha$ shown, the modes decay before the  onset of BBN, and the phenomenological limit disappear.

The ``scaling'' of the curves with $V_6^{1/6} / R$ shown in Figure~\ref {fig-long} can be also easily understood. From Eq.~(\ref{tdecalong}),  we see that 
the lifetime is a decreasing functions of $V_6^{1/6} / R $. This decrease is ``compensated'' by the growth of $\alpha$ in the  numerical prefactor
 (which is by far the most sensitive quantity on $ \alpha$ in that relation). Therefore, as $V_6^{1/6} / R$ increases, the  same lifetime and the same
 ``pattern'' in the phenomenological limit  is obtained by a small (logarithmic) increase of $\alpha \,$). We  see, in general, that for
 any given value of $\alpha$, the decay can  take place before BBN, provided the volume of the compact space is  sufficiently large.

\begin{figure}[h]
\centerline{
\includegraphics[width=0.35\textwidth,angle=-90]{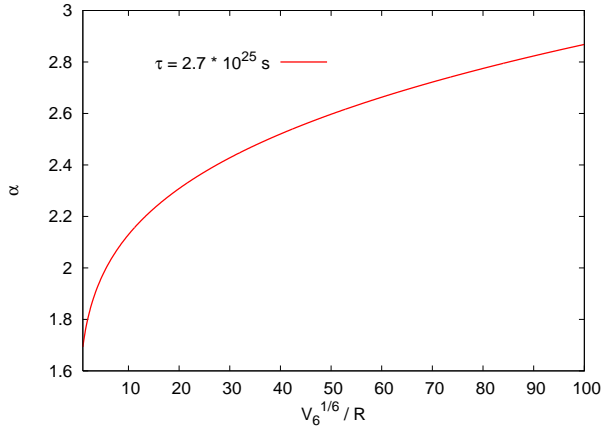}
}
\caption{Angular KK modes can be the dark matter candidate for $R/\sqrt{\alpha'} \simeq 6$ and  
for the values of the parameters $\alpha$ and $V_6^{1/6} / R$ above the line shown in  the plot, corresponding to a lifetime of the
 KK particles greater than about $3 \cdot 10^{25} \, {\rm s} \,$ (to avoid the limit from the diffuse $\gamma$ ray background,
 cf. Fig.~\ref{fig-limit}). For $V_6^{1/6} / R > 1 $ this occurs for relatively large values of $\alpha > 1.7$.}
\label{fig-dm}
\end{figure} 

We conclude by commenting on the case in which the angular KK modes are identified with the dark matter of the universe.
 As we have already mentioned, the right abundance is obtained for 
$R / \sqrt{\alpha'} \simeq 6 \,$. Moreover, limit from the diffuse $\gamma$ ray background require that
the lifetime of the modes is greater than about $3 \cdot 10^{25} \, {\rm s} \,$ (cf. Fig.~\ref{fig-limit}).
 The line in Figure~\ref{fig-dm} represents the point in the $V_6^{1/6} / R \,-\, \alpha$ plane corresponding to this lifetime.
 Points below this line give a shorter lifetime, and are therefore phenomenologically excluded. In particular we 
see that $\alpha > 1.7$ is required for any value of $V_6^{1/6} / R > 1$. Therefore, the reference value $\alpha = 1.29
$ does not allow to identify the angular KK modes with the dark matter of our universe.

\section{ Discussion and Conclusion}

In this paper, we have studied the problem of angular KK relics 
in string theory cosmology. As a specific example we considered angular KK modes 
resulting from
approximate isometries of the internal space in the context of warped flux
compactifications. We focused on  the case where the isometries of the warped throat are
distorted by its embedding into the compact manifold. The profile of the
different isometry breaking perturbations of the throat (the
parameters $\alpha_L$ in (\ref{series})) has been studied in \cite{Aharony:2005ez} for a
Klebanov-Strassler background, and we took their results as reference values. 

We investigated the interactions and possible decay channels of the KK modes with angular
momentum, calculating the corresponding coupling constants.
We did explicit calculations for the
spin 2 KK modes. We found that the lightest modes with angular momentum cannot
decay directly into any other two KK modes (massive or massless), independently
of the isometry breaking perturbations.
We considered the model where the SM particles are localized on a (probe) $3$-brane inside the throat. 
In this model the leading decay channel of the angular KK relics is the decay into  brane degrees of freedom. 
Different KK modes may decay through isometry breaking perturbations with different exponents $e^{-\alpha y_t / R}$, resulting 
in exponentially different lifetimes. 
Our final expression for the lifetime of the relic KK modes then depends
on several  parameters: the degree of isometry breaking at the tip of the
throat $e^{-\alpha y_t / R}$, the  volume of the internal space $V_6$,
the AdS curvature radius $R$, and the throat warping $e^{A}=e^{- y_t / R}$. For completeness we also included 
as a parameter the position of the SM brane inside the throat $e^{- y_b / R}$.
Remarkably, the coupling between angular KK modes and the SM particles can be much weaker than the
gravitational coupling.

We then studied the cosmological constraints on the parameters of the  single throat model.
We distinguished two limiting cases:  a short throat where brane inflation occurs, and
a long throat related to the  hierarchy problem.
In both cases, we considered the Standard Model fields to be located at the tip
of the throat and to be at some moment in thermal equilibrium with the lightest KK modes. 
The freeze-out abundance of the KK relics then depends essentially on a single
parameter $(R/\sqrt{\alpha'})$, although the dependance is very strong.

We considered the influence of the KK relics on the thermal history of the
universe and put together observational constrains on the parameters of KK modes and
 consequently, on  the underlying throat model, using limits from BBN \cite{moroi}
and the astrophysical $\gamma$-ray background \cite{diffusegamma}, in the case of decaying modes, and limits
on the dark matter energy density, in the case of lifetimes much greater than the present age of the universe.
Such limits are very sensitive to the parameters of the models (see Fig.~\ref{fig-bbnshort} for the short
throat, and Fig.~\ref{fig-long} for the long throat cases), but we can exhibit some general trends. 

In the short throat, the lifetime of the KK relics is much shorter than in the long throat (since the warp factor 
is smaller and, consequently, physical times are quicker), but their relic abundance is also much higher. 
In this case, we must require that these modes decay before the onset of BBN, namely with a lifetime 
$\tau \lta 10^{-2} \, {\rm s} \,$. For greater lifetimes, the BBN and overclosure limits
are typically exceeded by many 
orders of magnitude ; therefore, the demand that the decay takes place before the onset of BBN holds even if the thermal
 history for the KK modes is slightly different than the one considered here (for instance, even if they do not reach 
perfect thermal equilibrium before they decay).
We also showed that, for natural values of the parameters, the KK modes can dominate the energy density of
 the universe before decaying, see Fig.~\ref{fig-regshort}. This leads to an intermediate stage of matter
 domination in the early universe, which may have interesting consequences. For instance, it alters the
 thermal history of the universe, it 
  affects the relation between the  wavelength  of present-day cosmological fluctuations and the
 corresponding number of e-folds of inflation, etc.

 For the long throat case, a large fraction of the parameters space is excluded,
 see Fig.~\ref{fig-long}. 
In this case, for typical values of $\alpha$ obtained in \cite{Aharony:2005ez} (such as $\alpha=1.29$ for their leading 
isometry breaking perturbation), the constraints from the late decays require a sufficiently small value of $R/\sqrt{\alpha'}$ ($\lta 5$, or smaller),
 so that the abundance of the modes is strongly (Boltzmann) suppressed (recall that $R/\sqrt{\alpha'} \gta 1$ is required for 
the validity of the supergravity treatment of the throat geometry).
Otherwise, lower values of $\alpha$, or a sufficiently large compact space ($V_6^{1/6} / R \gta 100$), is required 
for these modes to decay before BBN. In the opposite regime, the lifetime of the KK relics can be much greater than 
the present age of the universe. If this is the case, or if, due to selections rules, some modes are stable, one gets 
the right abundance for these modes to be viable cold dark matter candidates if one tunes the value of 
$R/\sqrt{\alpha'} \simeq 6.1$.
This also requires a  value of $\alpha \gta 1.7$. 

In our  analysis, we considered the Standard Model degrees of freedom to be
located on a probe 3-brane at the tip of the throat where the KK modes are
localized. 
 The ultimate fate of the KK relics in a given model will depend on
the specific way that the Standard Model is realized and embedded into the
internal space. 
Furthermore, different or additional constraints may apply if
the Standard Model degrees of freedom are located in the unwarped region of the
Calabi-Yau or in another throat.

Already a displacement of the SM brane from the tip of the throat would affect the 
decay rate of the KK modes.
In Sections 5 and 6, we only considered the probe brane to be at the tip, $y_b = y_t$.
Otherwise $(y_b < y_t)$, the decay time of the relic KK modes into SM particles would be much bigger, 
see Eq.~(\ref{lamKKbbe}). Shifting the SM brane from the tip has qualitatively similar effects than  
an increase of  $\alpha$, but quantitatively it is more involved.
If $y_b < y_t$, all the KK modes interact much more weakly with the brane. In 
this case, even the modes with zero angular momentum can be dangerous relics, and observational 
constraints would be more stringent.

We restricted ourselves to the single throat case.
The reason is related to at least two additional complications that arise if we include
 another throat. One complication arises for two throats with very different warpings.
The physics of inflation and subsequent thermalization will corresponds to  values of the Hubble parameter and of
 the temperature which exceeds the mass of the KK modes in the long throat ($\sim TeV$).
This generates a mass gap for the KK modes. The tunneling of KK modes from the short to long throat
should take  this effect into consideration.
As it was recently shown in \cite{BK}, in such a scenario the Standard Model throat (together with SM brane) 
will be cloaked by a Schwarzschild horizon, and the usual treatment of tunneling is not applicable.

The second complication is even more restrictive. Let us try to model the wave equation of the 
 the angular KK modes in the bulk space between throats \cite{lev}. Since warping there is insignificant, instead of
equation (\ref{isophi}) we have 
\be
\label{bulk}
\left(\partial_y^2+m_{KK}^2-\frac{p^2}{R^2}\right)\Phi_m=0 \ ,
\ee
where we continue to use the ``radial'' coordinate $y$ (in a loose sense) and 
where $p^2$ is the eigenvalue of the Laplace operator $\Delta_f$ in the directions orthogonal to $y$.
Assuming that the angular momentum is connected to $p^2$, we conclude that
the angular KK modes are exponentially suppressed in the bulk,
$\Phi_m \sim e^{-p(L) |y| / R}$.
For $V_6^{1/6}/R \gg 1$ this effect exponentially suppresses the tunneling rate.  

 An interesting extension of the present analysis would be a more detailed study of the thermal history of the
 KK modes (or glueballs, in the  dual theory), from their production by the decay of heavy closed strings
 formed after brane annihilation, to their thermalization among themselves and with light SM fields. 
In this work, we have conservatively assumed that complete thermalization is quickly achieved; 
this reduces significantly the abundance of the KK modes (due to Boltzmann suppression). 
One should expect the KK modes to have a much higher abundance if thermalization is not complete;
 this would lead to even stronger bounds on the parameters of the model than those obtained here.

We note the paper \cite{artur}, which also made connection between cosmology and the glueballs in
the dual picture of the gauge theory with extra isometries, which appeared as the present paper was being completed.

\vskip 0.5in
\vbox{
\noindent{ {\bf Acknowledgments} } \\ 
\noindent
We thank Neil Barnaby, Alex Buchel, Igor Klebanov, Andrei Linde, Rob Myers,  Keith A. Olive, Joe Polchinki, Sergey Prokushkin,
 Mikhail Voloshin and Piljin Yi for  useful discussions. 
J.F.D. thanks CITA and the W.Fine Theoretical Physics Institute
and the School of Physics and Astronomy of the University of Minnesota
for their hospitality during different stages of this work.
The work of J.F.D. was  supported by CITA and by the Spanish MEC, via FPA2006-
05807 and FPA2006-05423. The work of M.P. was partially supported by DOE grant
DE--FG02--94ER--40823. L.K. was supported by NSERC and CIFAR. L.K. also  thanks KITP Santa Barbara for hospitality,
where this work was completed  during the Program Nonequilibrium Dynamics in Particle
Physics and Cosmology, and supported in part by the National Science Foundation under Grant No. PHY05-51164
}
\vskip 0.5in

\centerline{{\bf APPENDIX}}

In this Appendix, we derive the effective action to second and third order in 
the spin 2 KK modes (\ref{spin2}) around the general background (\ref{back}), assuming 
only 4-dimensional Lorentz invariance but independently of any isometries in the 
internal space. We then show that a massive mode cannot decay into one or two zero 
mode, generalizing a similar discussion in Section \ref{sec:interactions}. 

We consider the bulk action
\be 
\label{Sbulk}
S = \frac{M_D^{D-2}}{2} \int d^Dx\,\sqrt{-G}\,\left(R + 2\,\mathcal{L}\right) \ ,
\ee 
where the lagrangian $\mathcal{L}$ corresponds to all the matter in the bulk 
(scalar fields, forms, ...), as well as possible brane localized sources. If 
the spacetime has boundaries, we should also add the appropriate Gibbons-Hawking 
boundary terms. The ansatz for the $D$-dimensional metric is
\be
\label{metric}
ds^2 = G_{M N} dx^M dx^N = e^{2 A(y^c)}\,g_{\mu \nu}(x^\lambda, y^c)\,dx^\mu dx^\nu + \hat{g}_{a b}(y^c)\,dy^a dy^b \ ,
\ee
where capital latin indices $M, N, ...$ run over the $D$-dimensional coordinates, greek indices 
$\mu, \nu, \lambda, ...$ run over the 4-dimensional coordinates, and latin indices $a, b, c, ...$ 
run over the internal $(D-4)$-dimensional coordinates of the compact space. 

The background solution with 4-dimensional Lorentz symmetry corresponds to $g_{\mu \nu} = \eta_{\mu \nu}$ 
in (\ref{metric}). For the spin 2 perturbations, $g_{\mu\nu}$ will also 
depend on the coordinates $x^\mu$ and $y^c$, while the metric of the internal space 
$\hat{g}_{a b}$ depends only on the extra-dimensions coordinates $y^c$.

Einstein equations are
\be
\label{Eins}
R_{M N} = T_{M N} - \frac{T}{D-2}\,G_{M N} \;\;\;\; \mbox{ with } \;\;\;\; 
T_{M N} = \mathcal{L}\,G_{M N} - 2\,\frac{\delta \mathcal{L}}{\delta G^{M N}} \ ,
\ee
where $T$ is the trace of the energy-momentum tensor, $T = G^{M N} T_{M N}$. 
The background symmetries correspond to a matter lagrangian which is independent 
of the 4-dimensional metric: $\frac{\delta \mathcal{L}}{\delta g^{\mu \nu}} = 0$. The $(\mu, \nu)$ 
and $(a, b)$ components of Eq.~(\ref{Eins}) then read
\be
\label{einsmunu}
R_{\mu \nu} = \frac{2}{D-2}\,\left(-\mathcal{L} + 
\hat{g}^{c d}\,\frac{\delta \mathcal{L}}{\delta \hat{g}^{c d}}\right)\;e^{2A}\,g_{\mu \nu} \ ,
\ee
and
\be
\label{einsab} 
R_{a b} = \frac{2}{D-2}\,\left(-\mathcal{L} + 
\hat{g}^{c d}\,\frac{\delta \mathcal{L}}{\delta \hat{g}^{c d}}\right)\;\hat{g}_{a b} 
- 2\,\frac{\delta \mathcal{L}}{\delta \hat{g}^{c d}} \ .
\ee

It is straightforward to compute the Ricci tensor of the metric (\ref{metric}). 
The $(\mu, \nu)$ components are
\ba
\label{riccimunu}
R_{\mu\nu}[G] &=& R^{(4)}_{\mu\nu}[g] - \frac{1}{2}\,\hat{\nabla}_c \hat{\nabla}^c 
\left(e^{2A} g_{\mu\nu}\right) \nonumber \\
& + &
\frac{1}{4}\,e^{-2A}\,g^{\lambda \rho}\,
\left[2\,\partial_c\left(e^{2A} g_{\mu\rho}\right)
\,\partial^c\left(e^{2A} g_{\nu\lambda}\right) - 
\partial_c\left(e^{2A} g_{\lambda\rho}\right)\,\partial^c\left(e^{2A} g_{\mu\nu}\right)\right]
\ea
where $R^{(4)}_{\mu\nu}[g]$ is the $4$-dimensional Ricci tensor of the metric $g_{\mu\nu}$, 
$\hat{\nabla}_a$ is the covariant derivative associated with the metric $\hat{g}_{ab}$, 
and $a, b, c, ...$ indices are raised and lowered with the metric $\hat{g}_{a b}$. 
Similarly, the $(a, b)$ components of the $D$-dimensional Ricci tensor read
\be
\label{ricciab}
R_{a b}[G] = \hat{R}_{a b}[\hat{g}] - \frac{1}{2}\,\hat{\nabla}_a
\left[e^{-2A}g^{\mu \nu}\,\partial_b\left(e^{2A}g_{\mu \nu})\right)\right] - 
\frac{1}{4}\,e^{-4A}\,g^{\mu \nu}\,g^{\lambda \rho}\,\partial_a
\left(e^{2A}g_{\mu\lambda}\right)\,\partial_b\left(e^{2A}g_{\nu\rho}\right)
\ee 
where $\hat{R}_{a b}[\hat{g}]$ is the $(D-4)$-dimensional Ricci tensor of the metric 
$\hat{g}_{a b}$. To compute the effective action, we will also need the $D$-dimensional 
Ricci scalar for the ansatz (\ref{metric}). From Eqs. (\ref{riccimunu}, \ref{ricciab}), 
and performing the derivatives of the warp factor, we get
\ba
\label{ricciscal} 
R[G] &=& e^{-2A}\,R^{(4)}[g] + \hat{R}[\hat{g}] + 
\frac{1}{4}\,g^{\mu\nu}\,g^{\lambda \rho}\,
\left(\partial_c g_{\mu\rho}\,\partial^c g_{\nu\lambda} - 
\partial_c g_{\lambda\rho}\,\partial^c g_{\mu\nu}\right) - 
\frac{1}{2}\,\partial^c g^{\mu\nu}\,\partial_c g_{\mu\nu} - 
\nonumber \\ && 
g^{\mu\nu}\,\hat{\nabla}_c\hat{\nabla}^c g_{\mu\nu} 
- \partial^c A\,\left(6 g^{\mu\nu}\partial_c g_{\mu\nu} + 
g_{\mu\nu} \partial_c g^{\mu\nu}\right) - 8\,\hat{\nabla}_c\hat{\nabla}^c A - 
20\,\partial_c A\,\partial^c A
\ea
where $R^{(4)}[g]$ and $\hat{R}[\hat{g}]$ are respectively the $4$-dimensional and 
$(D-4)$-dimensional Ricci scalars of the metrics $g_{\mu\nu}$ and $\hat{g}_{a b}$.

We first consider the background solution, with $g_{\mu\nu} = \eta_{\mu\nu}$. 
Eq.~(\ref{einsmunu}) with (\ref{riccimunu}) then gives
\be
\label{warp}
\hat{\nabla}_c\hat{\nabla}^c A + 4\,\partial^c A\,\partial_c A = 
\frac{2}{D-2}\,\left(\mathcal{L} - 
\hat{g}^{c d}\,\frac{\delta \mathcal{L}}{\delta \hat{g}^{c d}}\right) \ .
\ee
Similarly, contracting Eq.~(\ref{einsab}) with $\hat{g}^{ab}$ and using 
Eq. (\ref{ricciab}) gives
\be
\label{RLA}
\hat{R}[\hat{g}] + 2\,\mathcal{L} = 6\,\hat{\nabla}_c\hat{\nabla}^c A 
+ 12\,\partial_c A\,\partial^c A \ ,
\ee
where we have used Eq.~(\ref{warp}) to eliminate the term in 
$\hat{g}^{c d} \frac{\delta \mathcal{L}}{\delta \hat{g}^{c d}}$. 

We now consider the spin 2 perturbations $h_{\mu\nu}(x,y)$, defined in (\ref{spin2}).
To obtain their effective action, we start from Eq.~(\ref{Sbulk}) with (\ref{ricciscal}) 
and $\sqrt{-G} = \sqrt{\hat{g}}\,\sqrt{-g}\,e^{4A}$. We may then eliminate the background 
$(D-4)$-dimensional Ricci tensor $\hat{R}[\hat{g}]$ and matter lagrangian $\mathcal{L}$ 
by using the relation (\ref{RLA}). This gives
\be
\label{S12}
S = S_1 + S_2
\ee
with
\be
\label{S1}
S_1 = \frac{M_D^{D-2}}{2} \int d^Dx\,\sqrt{\hat{g}}\,\sqrt{-g}\;e^{2A}\,R^{(4)}[g] \\
\ee
and
\ba
\label{S2}
S_2 &=& \frac{M_D^{D-2}}{2} \int d^Dx\,\sqrt{\hat{g}}\,\sqrt{-g}\;e^{4A}
\left[\frac{1}{4}\,g^{\mu\nu}\,g^{\lambda \rho}\,\left(\partial_c g_{\mu\rho}\,\partial^c g_{\nu\lambda} 
- \partial_c g_{\lambda\rho}\,\partial^c g_{\mu\nu}\right) 
- \frac{1}{2}\,\partial^c g^{\mu\nu}\,\partial_c g_{\mu\nu} - \right.  \nonumber \\
 &-& \left. g^{\mu\nu}\,\hat{\nabla}_c\hat{\nabla}^c g_{\mu\nu} - \partial^c A\,\left(6 g^{\mu\nu}\partial_c g_{\mu\nu} + g_{\mu\nu} \partial_c g^{\mu\nu}\right) 
- 2\,\hat{\nabla}_c\hat{\nabla}^c A - 8\,\partial_c A\,\partial^c A\right] \ .
\ea
 
We first expand the action to quadratic order in $h_{\mu\nu}$. After some integrations 
by parts (and neglecting the boundary terms), we can cast the action in the form
\be
\label{Squad}
S^{(2)} = \frac{M_D^{D-2}}{8} \int d^Dx\,\sqrt{\hat{g}}\,e^{2A}\, 
h^{\mu \nu} \left[\Box_{(4)} + e^{2A}\,\left(\hat{\nabla}_c\hat{\nabla}^c h_{\mu\nu} 
+ 4\,\partial^c A\,\partial_c \right)\right] h_{\mu\nu} \ ,
\ee
where greek indices are now raised and lowered with the $4$-dimensional Minkowski metric 
$\eta_{\mu\nu}$ and $\Box_{(4)} = \eta^{\mu\nu} \partial_\mu \partial\nu$. The vanishing 
of the terms inside the brackets gives the equations of motion to first order in 
$h_{\mu\nu}$. With the separation of variables (\ref{sepvar}), this gives Eq. 
(\ref{equaphi}) for the profile of the modes in the internal space. 
The action~(\ref{Squad}) then becomes
\begin{equation}
S^{(2)} = \frac{M_D^{D-2}}{8} \sum_{m,m'}  \int d^{D-4} y \sqrt{\hat g} \, e^{2 A} 
\Phi_m \Phi_{m'} \int d^4 x \, \gamma^{\mu \nu \, (m')} \left[ \Box_{(4)} - m^2 \right] 
\gamma_{\mu \nu}^{(m)} \ .
\end{equation}
Therefore, with the orthonormal conditions
\begin{equation}
\frac{M_D^{D-2}}{4} \int d^{D-4} y \sqrt{\hat g} \, e^{2 A} \Phi_m \Phi_{m'} 
= \delta_{m m'} \, ,
\label{orthonorm}
\end{equation}
the quadratic action decouples, and each $\gamma_{\mu \nu}^{(m)}$ represents from the 
4-dimensional point of view a canonically normalized spin $2$ field of mass $m$. For 
self-adjoint boundary conditions, the orthogonality conditions in (\ref{orthonorm}) 
follows from the mode equation (\ref{equaphi}) in the usual way, by multiplying 
(\ref{equaphi}) by $\Phi_{m'}$, integrating by part over the internal space, and subtracting 
the equation obtained by interchanging $m$ and $m'$. 

Eq.~(\ref{equaphi}) admits in particular a massless mode with a constant wave function 
in the internal space, $\Phi_0 = \mathrm{constant}$. This mode represents the standard 
graviton in the 4-dimensional picture. The effective 4-dimensional Planck mass may then 
be obtained by performing the $y$-integral in (\ref{S1}) for a 4-dimensional metric 
$g_{\mu\nu}(x)$ depending only on the external coordinates
\begin{equation}
\Mp^2 = M_D^{D-2} \int d^{D-4} y \sqrt{\hat g} \, e^{2 A} = M_D^{D-2} \, V_{D - 4} \ ,
\label{mp}
\end{equation}
where $V_{D - 4}$ is the volume of the internal space. Using the normalisation 
(\ref{orthonorm}), the wave function of the zero mode is then
\be
\label{graviton}
\Phi_0 = \frac{2}{\Mp} \ .
\ee

We now expand the action to third order in $h_{\mu\nu}$, in order to obtain the trilinear 
interactions between the spin 2 KK modes. For concreteness, we consider the decay of a mode 
of mass $m$ into two modes of mass $m'$ and $m''$. The first term in (\ref{S12}, \ref{S1}) 
gives contributions of the form 
\be
\label{S3}
S_1^{(3)} \supset 
\frac{M_D^{D-2}}{2} \int d^{D-4}y\,\sqrt{\hat{g}}\,e^{2A}\,\Phi_m\,\Phi_{m'}\,\Phi_{m''} \; 
\int d^4x\,\gamma^{(m)\,\nu}_\mu \partial_\sigma 
\gamma^{(m')}_{\nu\rho} \partial^\sigma \gamma^{(m'')\,\mu\rho} \ ..
\ee
In general, the $y$-integrals do {\em not} require the conservation of the total KK mass, 
$m = m' + m''$, as would have been the case in a flat background. However, if one of the 
decay product is the zero-mode, $m'' = 0$, its wave-function is constant and goes out of 
the integral. The $y$-integrals then reduce to the orthonormal conditions (\ref{orthonorm}), 
which imposes $m = m'$. Therefore, a massive KK mode ($m \neq 0$) cannot decay into 2 zero 
modes gravitons ($m = m' = m''$), as also noticed in \cite{tye2}. Here we note furthermore 
that a massive mode cannot decay either into another massive mode plus a graviton since, 
for the process $m \rightarrow m' + 0$, the constraint $m = m'$ is not compatible with the 
conservation of the 4-dimensional energy-momentum (\ref{Econs}). We thus conclude that a 
massive KK mode cannot decay directly by 3-legs interaction into one or two zero-mode(s).

We can check that the second term in (\ref{S12}) does not change these conclusions. 
Expanding (\ref{S2}) to third order in $h_{\mu\nu}$ and performing integrations by part, 
we find after a tedious calculation
\ba
\label{S23}
S_2^{(3)} &=& \frac{M_D^{D-2}}{2} \int d^Dx\,\sqrt{\hat{g}}\,e^{4A}\,
\left[\frac{1}{2}\,h^{\mu\nu}\,\partial_c h_{\nu\lambda}\,\partial^c h^\lambda_\mu 
- \partial^c A \, h^{\mu\nu} \, h_{\nu\lambda}\, \partial_c h^\lambda_\mu - \right. \nonumber \\
&-& \left. \frac{1}{3}\,\left(\hat{\nabla}_c\hat{\nabla}^c A 
+ 4 \,\partial_c A\,\partial^c A\right)\,h^{\mu\nu} \, h_{\nu\lambda}\,h^\lambda_\mu \right] \ .
\ea
We are interested in the contribution of this expression to the process 
$m \rightarrow m' + 0$. There is only one possibility for the constant zero-mode in the 
first term, 2 in the second term and 3 in the third term. The 4-dimensional coupling 
constant for this process is then proportional to
\ba
M_D^{D-2}\Phi_0\,\int d^{D-4}y\,\sqrt{\hat{g}}\,e^{4A}
\left[\frac{1}{2}\,\partial_c\Phi_m\,\partial^c\Phi_{m'} 
-\partial^c A\,\partial_c \left(\Phi_m\,\Phi_{m'}\right) \right. \nonumber \\
- \left. \left(\hat{\nabla}_c\hat{\nabla}^c A + 
4 \,\partial_c A\,\partial^c A\right)\,\Phi_{m'}\Phi_m\right] \ .
\ea
By integrating the first and second terms by part, and by using the mode equation 
(\ref{equaphi}), this reduces to
\be
\frac{M_D^{D-2}}{4}\,\Phi_0\,\left(m^2 + m^{,\,2}\right)\,
\int d^{D-4}y\,\sqrt{\hat{g}}\,e^{2A}\,\Phi_m\,\Phi_{m'} \ .
\ee
which again is non-vanishing only for $m = m'$, from the orthogonal conditions 
(\ref{orthonorm}). 

Therefore, the 3-legs decay of a massive spin 2 KK mode into 1 or 2 graviton zero-mode(s) 
is not possible. We will use this in Section \ref{sec:interactions} when we study the possible 
decay channels of the KK modes.\\



\end{document}